\documentclass[11pt]{article}

\pdfoutput=1 
\usepackage[pdftex]{graphicx}
\usepackage{booktabs}
\usepackage[pdftex]{xcolor}

\usepackage[margin = 1in]{geometry}
\usepackage{array, amsmath, graphicx, mathtools, amssymb, amsfonts, mathtools, setspace, booktabs, float, caption, pgfplots}
\usepackage[title]{appendix}
\usepackage{palatino}
\usepackage[noblocks]{authblk}
\usepackage{subcaption, caption, abstract, txfonts, longtable, adjustbox}

\graphicspath{{Figures/}}

\usepackage{url}

\DeclareCaptionFormat{cont}{#1 (cont.)#2#3\par}

\usepackage{mathpazo, wrapfig,  tikz, enumitem} 

\usepackage[scaled]{helvet} 
\usepackage{courier} 
\normalfont
\usepackage[T1]{fontenc}
\usepackage[utf8]{inputenc}

\graphicspath{{FIGURES/}}

\newcommand{\length}{0.7\textwidth}

\def\Myitem[#1]#2{\item[{\rm #1}]#2}

\providecommand{\keywords}[1]
{
  \small	
  \textbf{\textit{Keywords---}} #1
}

\begin{document}

\title{A comparison of partial information decompositions using data from real and simulated layer 5b pyramidal cells}
\author[1]{Jim W. Kay} 
\author[2]{Jan M. Schulz}
\author[3]{William A. Phillips}

\affil[1]{School of Mathematics and Statistics, University of Glasgow, UK, jim.kay@glasgow.ac.uk}
\affil[2]{Department of Biomedicine, University of Basel, Switzerland, j.schulz@unibas.ch}
\affil[3]{Faculty of Natural Sciences, University of Stirling, UK, w.a.phillips@stirling.ac.uk}

\maketitle

\begin{abstract}
Partial information decomposition allows the joint mutual information between an output and a set of inputs to be divided into components that are synergistic or shared or unique to each input. We consider five different decompositions and compare their results on data from layer 5b pyramidal cells in two different studies. The first study was of the amplification of somatic action potential output by apical dendritic input and its regulation by dendritic inhibition. We find that two of the decompositions produce much larger estimates of synergy and shared information than the others, as well as large levels of unique misinformation. When within-neuron differences in the components are examined, the five methods produce more similar results for all but the shared information component, for which two methods produce a different statistical conclusion from the others. There are some differences in the expression of  unique information asymmetry among the methods. It  is  significantly larger, on average, under dendritic inhibition. Three of the methods support a previous conclusion that apical amplification is reduced by dendritic inhibition.
The second study used a detailed compartmental model to produce action potentials for many combinations of the numbers of basal and apical synaptic inputs. Decompositions of the entire data set produce similar differences to those in the first study. Two analyses of decompositions are conducted on subsets of the data. In the first, the decompositions reveal a bifurcation in unique information asymmetry. For three of the methods this suggests that apical drive switches to basal drive as the strength of the basal input increases, while the other two show changing mixtures of information and misinformation.  Decompositions produced using the second set of subsets show that all five decompositions provide support for properties of cooperative context-sensitivity - to varying extents.
\end{abstract}


\section{Introduction}

A breakthrough in information theory happened in 2010 when Williams and Beer~\cite{WB} published a method called {\it partial information decomposition} which provided a framework under which the mutual information, shared between the inputs and the output of a probabilistic system, could be decomposed into components which measure different aspects of the information: the unique information that each input conveys about the output; the shared information that all inputs possess regarding the output; the information that the inputs in combination have about the output. They also defined  a decomposition method, Imin. Several authors criticised the definition of the  redundancy component in the Imin method~\cite{HSP, BERT, GK, RI} thus spawning  several new methods for computing a partial information decomposition (PID), including: Ibroja which was developed independently by Bertschinger, Rauh, Olbrich, Jost and Ay~\cite{BERT}  and by Griffiths and Koch~\cite{GK}, Idep invented by James, Emenheiser and Crutchfield~\cite{James}, Iccs invented by Ince~\cite{RI}, Ipm a PID invented by Finn and Lizier~\cite{FL}, and Isx developed by Makkeh, Gutknecht and Wibral~\cite{MGW, IDTxl}. The Ibroja and Idep PIDs are guaranteed to have nonnegative components while the other three methods are defined in a pointwise manner by considering information measures at the level of individual realisations and defining partial information components at this level. Pointwise PIDs can produce negative components and they are described as providing {\it misinformation} in such cases~\cite{locinf} 

An important feature of PID is that it enables the shared information and synergistic information in a system to be estimated separately. This provides an advance on earlier research in which the interaction information was used to estimate synergy~\cite{SBB, GT}, but could be negative, and the three-way mutual information~\cite{KFP} or coinformation~\cite{Bell}, which also could be negative, was used as an objective function in neural networks with two distinct sets of inputs from receptive and contextual fields, respectively. 

Partial information decomposition has been applied to data in neuroscience, neuroimaging, neural networks and cellular automata; see, e.g.~\cite{Wibral} --\cite{WLP}. A major selling point of the Isx PID is that the components are differentiable, unlike other PIDs. This makes it possible to build neural networks with a particular neural goal involving PID components  as the objective function~\cite{WPKLP, MG}. For an overview of PID, see~\cite{LBJW}, and for an excellent tutorial, see~\cite{TL}.

We will provide a systematic comparison of the different methods by applying the Ibroja, Idep, Iccs, Ipm and Isx PIDs to  data recorded in two different studies. First, we present  more detailed analyses and comparisons of physiological data recorded from cortical layer 5b (L5b) pyramidal neurons in  a study of GABA$_{\rm B}$ receptor-mediated regulation of dendro-somatic synergy. The influence of GABA$_{\rm B}$ receptor-mediated inhibition of the apical dendrites evoked by local application of the GABA$_{\rm B}$ receptor agonist baclofen  will be studied by making within-neuron paired comparisons of PID components. We will also shed light on unique information asymmetries as revealed by the PID analyses, as well as discussing the evidence for apical amplification in the presence and absence of GABA$_{\rm B}$ receptor-mediated inhibition of apical dendrites.

The stereotypical morphology of pyramidal neurons suggests that they have at least two functionally distinct sets of fine dendrites, the basal dendrites that feed directly into the cell body, or soma, from where output action potential(AP)s are generated, and the dendrites of the apical tuft, which are far more distant from the soma and connected to it by the apical trunk. Inputs to the branches of the apical tuft arise from diverse sources that specify various aspects of the context within which the feedforward input to the basal dendrites is processed~\cite{RM, Schu}. These apical inputs are summed at an integration zone near the top of the apical trunk, which, when sufficiently activated, generates calcium-dependent regenerative potentials in the apical trunk, thus providing a cellular mechanism by which these pyramidal cells can respond more strongly to activation of their basal dendrites when that coincides with activation of their apical dendrites~\cite{ML2013}. Though these experiments require an exceptionally high level of technical expertise, there are now many anatomical and physiological studies indicating that some classes of pyramidal cell can operate as context-sensitive two-point processors, rather than as integrate-and-fire point processors~\cite{WS, ML2013, BNSPS}. Thick-tufted L5b pyramidal cells are the class of pyramidal cell in which operational modes approximating context-sensitive two-point processing have most clearly been demonstrated, but it may apply to some other classes of pyramidal cell also, though not to all~\cite{FW}. These advances in our knowledge of the division of labor between apical and basal dendrites now give the interaction between apical and basal dendritic compartments a prominent role within the broader field of dendritic computation~\cite{PP}.

The second study~\cite{Shai} considers data on spike counts obtained using an amended version of the Hay compartmental model~\cite{HHSMS}. Spike counts are available for many different combinations of basal and apical inputs. While PIDs can be computed for the entire dataset, an interesting diversity of balance between basal drive and apical drive is revealed by applying the PID methods to  subsets of the data defined by various combinations of basal and apical inputs. For one set of subsets this reveals a bifurcation in unique information asymmetry, and a difference among the methods in how this is expressed.  A different  analysis of subsets allows a full discussion of the extent to which evidence of  cooperative context-sensitivity  is revealed by the nature of the PID components.

Many empirical findings have been interpreted as indicating that cooperative context-sensitivity  is common throughout perceptual and higher cognitive regions of the mammalian neocortex. For example, consider the effect of a flanking context on the ability to detect a short faint line. A surrounding context is neither necessary nor sufficient for that task, but many psychophysical and physiological studies show that context can have large effects, nevertheless, as reviewed, for example, by Lamme~\cite{Lamme1, Lamme2} and by~\cite{GS}.  In~\cite{KIDP, KP2020}, theoretical studies on  the effects of context (then called `contextual modulation') were explored. Here we have the opportunity to explore this topic using data from real and simulated L5b pyramidal cells. The ideal properties of cooperative context-sensitivity are described in the Methods section.

\section{Methods}

\subsection{Data}
Physiological data recorded from rat L5b pyramidal neurons during dual patch-clamp recordings from soma and apical dendrite before and during local application of the GABA$_{\rm B}$B receptor agonist baclofen was taken from~\cite{SKBL}. Spike count data obtained using an amended version of the Hay compartmental model was taken from~\cite{Shai}.
\subsection{Notation and Definitions}

We consider trivariate probabilistic systems involving three discrete random variables: an output $Y$ and two inputs $B$ and $A$. Hence, underlying the discrete data sets we consider is a probability mass function $\Pr(Y =y, B =b, A=a)$, where $y, b, a$ belong to the finite alphabets $\mathcal{A}_Y, \mathcal{A}_B, \mathcal{A}_A$, respectively. 

We now define the standard information theoretic terms that are required in this work and they are based on results in~\cite{CT}. We denote by the function $H$ the usual Shannon entropy, and note that any term with zero probability makes no contribution to the sums involved.
The joint mutual information that is shared by $Y$ and the pair $(B, A)$ is given by,
\begin{equation}
I(Y; B, A) = H(Y) + H(B, A) - H(Y, B, A). \label{totmi}
\end{equation}

The information that is shared between $Y$ and $B$ but not with $A$ is 
\begin{equation}
I(Y; B |A) = H(Y, A) + H(B, A) - H(A) - H(Y, B, A), \label{yx1Gx2}
\end{equation}
and the information that is shared between $Y$ and $A$ but not with $R$ is
\begin{equation}
I(Y; A |B) = H(Y, B) + H(B, A) - H(B) - H(Y, B, A). \label{yx2Gx1}
\end{equation}
The information shared between $Y$ and $B$ is
\begin{equation}
I(Y; B) = H(Y) + H(B) - H(Y, B)
\end{equation}
and between $Y$ and $A$ is 
\begin{equation}
I(Y; A) = H(Y) + H(A) - H(Y, A)
\end{equation}
The interaction information~\cite{McG} is a measure of information involving all three variables, $Y, B, A$ and is defined by
\begin{equation}
II(Y; B; A) = I(Y; B, A) - I(Y; B) - I(Y; A)  \label{IntInf}
\end{equation}

\subsection{Partial Information Decomposition}

 The information decomposition can be expressed as~\cite{WPKLP}
\begin{equation} I(Y; B, A) = I_{unq}(Y; B|A) +  I_{unq}(Y; A|B) + I_{shd}(Y; B, A) + I_{syn}(Y; B, A). \label{totalDecomp} \end{equation}

Adapting the notation of~\cite{WPKLP} we express our joint input mutual information in four terms as follows:

\begin{figure}[H]
\centering
\begin{center}
\begin{tabular}{p{0.3\textwidth} p{0.6\textwidth}}
$\text{Unq}B  \equiv  I_{unq}(Y; B|A)$ & denotes the unique information that $B$ conveys about $Y$;\\\\
$ \text{Unq}A  \equiv I_{unq}(Y; A| B)$ & is the unique information that $A$ conveys about $Y$;\\\\
 $\text{Shd~~~~} \equiv I_{shd}(Y; B, A)$ & gives the shared (or redundant) information that both $B$ and $A$  have about $Y$;\\\\
$\text{Syn~~~~} \equiv I_{syn}(Y; B, A)$ & is the synergy or information that the joint variable $(B, A)$ has about $Y$ that cannot be obtained by observing $B$ and $A$ separately. 
\end{tabular}
\end{center}

\end{figure}

It is possible to make deductions about a PID by using the following four equations which give a link between the components of a PID and certain classical Shannon measures of mutual information. The following are in~\cite[eqs. 4, 5]{WPKLP} , with amended notation; see also~\cite{WB}.
\begin{align}
I(Y; B) &=  \text{Unq}B +  \text{Shd} \label{ux1red}\\
I(Y; A) &=  \text{Unq}A +    \text{Shd}, \label{ux2red} \\
I(Y; B| A) &= \text{Unq}B +  \text{Syn}, \label{ux1syn}\\
I(Y; A | B) & =  \text{Unq}A + \text{Syn}. \label{ux2syn}
\end{align}

Using~$\eqref{totalDecomp}, \eqref{ux1red}, \eqref{ux2red}$ we may deduce the following connections between classical information measures and partial information components.
 \begin{equation}
 II(Y;B;A) = \text{Syn} - \text{Shd} \label{synshd}
 \end{equation}
 \begin{equation}
 I(Y; B) - I(Y; A) = \text{UnqB} - \text{UnqA} \label{unqs}
 \end{equation}

When the partial information components are known  {\it a priori}  to be non-negative, we may deduce the following  from~$\eqref{totmi}$, $\eqref{ux1red}$, $\eqref{ux2red}$. When the interaction information in~$\eqref{IntInf}$ is positive, a lower bound on the synergy of a system is given by the interaction information~\cite{McG}. Also, the expression in~$\eqref{unqs}$ provides a lower bound for UnqB, when $I(Y; B) > I(Y;A)$. Thus some deductions can be made without considering a PID. While such deductions can be useful in providing information  bounds, it is only by computing a PID that the actual values of the partial information components can be obtained. 

When making comparisons between different systems it is sometimes necessary to normalise the PID components by dividing each term by their sum, the joint mutual information, $I(Y; B, A)$. Such normalisation will be  applied in the analyses considered in the sequel. This means that the sum of the PID components is equal to unity and so they are negatively correlated.

In this study, the PID component, Shd, has not been separated into a sum of source, ShdS,  and mechanistic, ShdM, terms~\cite{HSP,  KIDP, PICA} as
\begin{equation*}
\text{Shd} = \text{ShdS} + \text{ShdM}
\end{equation*}
because not all of the five PIDs considered  include definitions regarding how to achieve this task. For probability distributions in which the inputs $B$ and $A$ are marginally independent the source shared information, ShdS, should be equal to zero, and hence the shared information, Shd, is entirely mechanistic shared information - shared information due to the probabilisitic mechanism involved in the information processing.

\subsection{Unique Information Asymmetry}
We define the unique information asymmetry (UIA) to be UnqB - UnqA. From $\eqref{ux1syn}, \eqref{ux2syn}, \eqref{unqs}$ we have that
\begin{equation}
\text{UnqB} - \text{UnqA}  = I(Y; B) - I(Y; A) = I(Y;A|B) - I(Y;B|A) \label{uia}
\end{equation}
The value of UIA is the same for every PID method. When UIA > 0, we say that the basal input is mainly driving, whereas when UIA < 0 it is the apical input that is mainly providing the drive. Asymmetries for which UIA > 0 and UnqA is zero or small in magnitude are of interest in relation to property CSS3 of cooperative context-sensitivity, as defined below.

\subsection{Pointwise PID methods}
The PID methods Ibroja and Idep produce PID components that are non-negative, whereas Iccs, Ipm and Isx can produce negative values. The PIDs Iccs, Ipm and Isx are pointwise-based methods in which local information measures are employed at the level of individual realizations of the random variables.   Local mutual information is explained by Lizier in~\cite{locinf}. If $U, V$ are discrete random variables then the mutual information $I(U, V)$ shared between $U$ and $V$ can be written as an average of the local mutual information terms $i(u; v)$, for each individual  realization $(u, v)$ of $(U, V)$, as follows
\begin{equation} 
I(U; V) =  \sum_{u, v} p(u, v)  \log_{2} \frac{p(u, v)}{p(u) p(v)} = \sum_{u, v} p(u, v)  \log_{2} \frac{p(u | v)}{p(u)} = \sum_{u, v} p(u, v)  \log_{2} i(u; v), 
\label{midecomp}\end{equation}
where 
\begin{equation*}
i(u; v) =  \log_{2} \frac{p(u | v)}{p(u)}  \end{equation*} is the local mutual information associated with the  realization $(u, v)$ of $(U, V)$.

The local mutual information $i(u; v)$ is positive when $p(u|v) > p(u)$, so that ``knowing the value of $v$ increased our expectation of (or positively informed us about) the value of the measurement $u$''~\cite{locinf}. The local mutual information $i(u; v)$ is negative when $p(u|v) < p(u)$, so that ``knowing about the value of $v$ actually changed our belief $p(u)$ about the probability of occurrence of the outcome $u$ to a smaller value $p(u | v)$, and hence we considered it less likely that $u$ would occur when knowing $v$ than when not knowing $v$, in a case were $u$ nevertheless occurred''~\cite{locinf}. Of course, the average of  these local measures is  the mutual information $I(U; V)$, as in~$\eqref{midecomp}$, but when pointwise information measures are used to construct a PID there can be negative averages. For further details of how negative values of PID components can occur, see~\cite{ RI,  FL, MGW, KIDP}. 

In the analyses reported below, it will be found  that the  pointwise  PIDs can  give negative values for the unique information due to $A$, or for the unique information due to $B$, or both.  We interpret this to mean that the unique information provided by $A$, or by $B$ is, on average,  less likely to result in predicting the correct value of the output $Y$.  We adopt  the term `misinformation' from~\cite{FL, MGW, locinf}, and describe this as 'unique misinformation due to $A$ (or $B$)'. 

\subsection{Ideal properties of cooperative context-sensitivity}
We now state key properties of cooperative context-sensitivity (which are a modified form of those specified for contextual modulation in~\cite{KP2020}), while recognising that in any biological system these properties are likely to be observed only as an approximation to the ideal. It is assumed that the basal input is driving and the apical input provides the context. The context amplifies the transmission of information about the necessary, or driving, input when criterion CCS3 is met.

\begin{enumerate}[leftmargin = 1.5cm]
\item[CCS1:] The drive, $B$, is sufficient for the output to transmit information about the input, so context, $A$, is not necessary.
\item[CCS2:] The drive, $B$, is necessary for the output to transmit information about the input, so context, $A$, is not sufficient.
\item[CCS3:] The output transmits unique information about the drive, $B$, but little or no unique information or misinformation about the context, $A$, although synergistic or shared mechanistic components, or both,  are present.
\item[CCS4:] The context strengthens the transmission of information about $B$ when $B$ is weak. As the strength of $B$ increases the synergy and shared mechanistic information decrease.
\end{enumerate}

\subsection{Statistics}

Summary statistics are presented as the sample median and the sample quartiles. Significance testing based on  within-neuron differences  is conducted using a two-sided exact Wilcoxon signed rank test of equality of population medians, and the threshold for declaring statistical significance of a single test is  P <  0.05. Where multiple tests are used the individual P values were corrected by using the Bonferroni method to ensure that the family-wise error rate is at most 0.05; if $m$ simultaneous tests are conducted, a test with p value P has a corrected value of $\min(m P, 1)$.

\subsection{Software}
The Ibroja PID was estimated using {\sf compute UI}~\cite{ADM}. The discrete information theory library {\sf dit}~\cite{dit} was used to estimate the  Iccs, Idep and Ipm PIDs. R~\cite{R} code was also used to estimate the Iccs and Idep PIDs. Python code was called from RStudio~\cite{R} by using the {\sf reticulate} package~\cite{reticulate}. The graphics were produced by using the {\sf ggplot2} package~\cite{ggplot2} in RStudio. Statistical testing made use of the {\sf coin} package~\cite{coin} in RStudio. 

\section{Results}

\subsection{Real data from patch-clamp recordings in L5b pyramidal neurons}
In a study of GABA$_B$ receptor-mediated regulation of dendro-somatic synergy in  L5b pyramidal neurons~\cite{SKBL},  the relationship between AP  output, $Y$,  to input currents during combined current injections into the soma, $S$ (referred to as basal input, $B$, in the sequel) and distant apical dendrite of thick-tufted L5b pyramidal neurons, $D$ (referred to as apical input, $A$, in the sequel) in rat somatosensory cortex was recorded. Current waveforms injected via patch-clamp pipettes were used to mimic synaptic responses to contralateral hind limb stimulation in vivo~\cite{Palmer}. AP trains were recorded for $\geq 25$ (range: 25-49) combinations of different current levels (see Figure 1). 

The normalised injected waveforms were scaled by separate amplification factors ranging from 0 pA   up to 1500 pA, resulting in at most 49 trials for each neuron.
Trials for which there were no APs for a treatment condition were omitted from consideration. Care was taken to ensure that the input distributions for the treatment conditions considered within a neuron contained exactly the same combinations of somatic and dendritic amplitude. This is particularly important since there is interest in comparing the PIDs obtained under different treatment conditions within each neuron. Ensuring that
the input (B, A) distributions match ensures that any observed difference in a PID component within a neuron is not simply due to a difference in the input distributions.
Data of time-varying input currents and resulting AP times, from the admitted trials,  were binned into non-overlapping segments of 120 ms to maximize the joint mutual information (see~\cite{SKBL}, Figure S1). Within each bin the AP number, the mean somatic and mean dendritic signals were computed. The values of each of the input signals were binned into quartiles to maximize entropy~\cite{TL}. The output was categorized as 0, 1 or 2+ APs. Thus, we generated a 4 by 4 by 3 probability distribution for each of the neurons considered under each of the treatment conditions.

\begin{figure}[H]
\centering 
\includegraphics[width= 12cm]{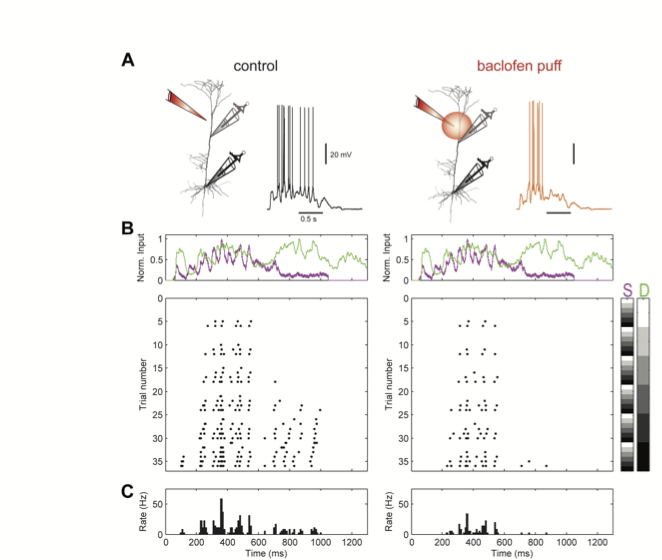} 

\caption{Dual dendritic and somatic patch-clamp recording from a L5b pyramidal neuron of rat somatosensory cortex enables the study of amplification of somatic AP output by apical dendritic input and its regulation by dendritic inhibition. (A) Locations of dual dendritic and somatic patch-clamp recordings are indicated on a biocytin-filled L5 pyramidal neuron. After recordings in the control condition, the GABA$_{\rm B}$ agonist baclofen (50 $\mu$M) was puffed onto the apical dendrite at 50 to 100 $\mu$m distal to the dendritic patch pipette. Example membrane potential responses to combined current injections into soma and dendrite are shown in control condition (left) and during the puff of baclofen (right). Peak current amplitude was 1,000 pA for dendritic and somatic current injections. (B) Top, injected current waveforms based on {\em in vivo} responses to sensory stimulation~\cite{Palmer}. Dendritic current is shown in green, somatic in purple. Bottom, raster plot of APs emitted in individual episodes during increasing levels of dendritic and somatic stimulation strength. Control is shown on the left. A raster plot of APs emitted in the same neuron during activation of dendritic GABA$_{\rm B}$Rs by a puff of baclofen onto the apical dendrite is shown on the right. Different levels of the injected current in 36 combinations are indicated by the right colour bars (S, somatic; D, dendritic). The peak amplitude of the current waveform was increased from 0 pA (white) to 1,250 (black) in soma and dendrite, respectively. Step size was 250 pA. (C) Peri-stimulus time histogram of APs across all current combinations for both conditions. All data was taken from~\cite{SKBL}.}
\label{fig2}
\end{figure}

\begin{figure}[H]
\centering
\includegraphics[width = 14 cm]{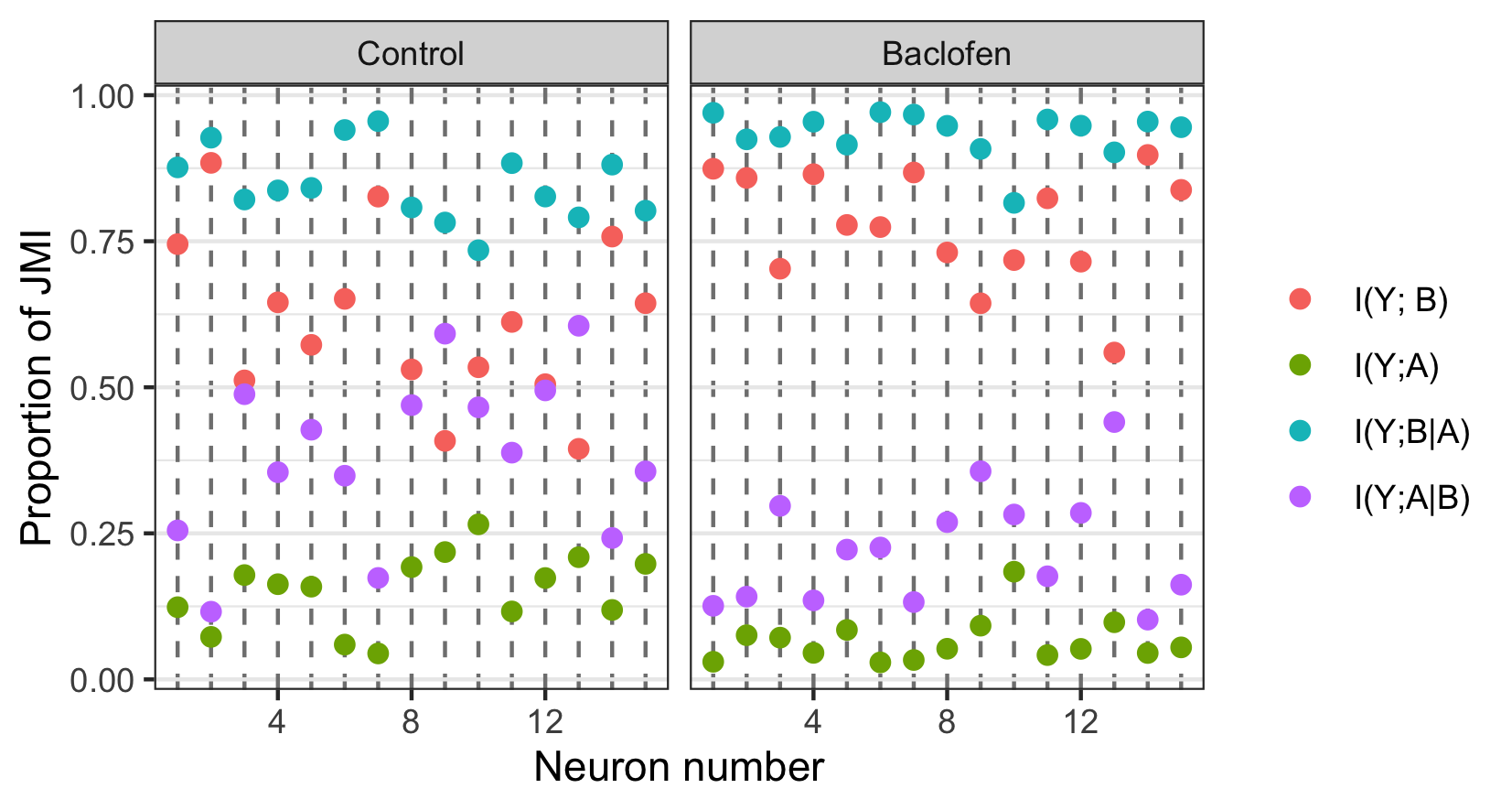}
\caption{Physiological L5b neuronal recording data. For each of the 15 neurons under  each of the Control and Baclofen  conditions, the values of the four normalised  classic mutual information measures that are involved in the definition of a PID are displayed. Their values are given as their relative contributions to the joint mutual information in each case.}
  \label{fig2}
\end{figure}

\subsubsection{Classic mutual information measures}

The classic mutual  information measures were computed for each neuron under each of the two experimental conditions. The values of the joint mutual information (JMI) between the AP count ($Y$) and the pair of basal and apical inputs ($B, A$) ranged from 0.49 to 0.93 bit for neurons in the Control condition and from 0.45 to 1.02 bit for neurons exposed to baclofen. Therefore the information measures computed for each neuron under each condition were normalised by dividing by their respective JMI values. Normalised values are displayed in Figure~\ref{fig2}.  
It is worth noting that when normalised the information measures values satisfy the equations
\begin{equation} I(Y;B)+ I(Y;A|B) = 1 = I(Y;A) + I(Y;B|A) \label{norm} \end{equation}
which means that $I(Y;B)$ and $I(Y;A|B)$ are negatively correlated, as are $I(Y;A)$ and $I(Y;B|A)$. 

Following the vertical dashed lines in Figure~\ref{fig2}, we notice that the mutual information between the AP count and the basal input is much larger than that between the AP count and the apical input, indicating clear unique information asymmetry. For these neurons the AP count is more strongly related to the basal than the apical input. This is the case for both of the experimental conditions.  We can therefore anticipate that each of the PIDs considered will also express these asymmetries between the values of their unique basal and apical PID components.

In Figure~\ref{fig2}, for most neurons under both conditions we see that when $I(Y;B|A)$ is high the corresponding $I(Y;A)$ values are low and when the $I(Y;B)$ values are large the corresponding $I(Y;A|B)$ values are small. These simply reflect the negative correlations between the respective normalised measures as a result of $\eqref{norm}$.

\begin{table}[H]
 \centering
 \caption{Physiological L5b neuronal recording data. Summary statistics for the sample of 15 neurons. 
Summary statistics of the normalised information measures, showing for each measure the median and interquartile range for the sample of 15 neurons under both experimental conditions, given as percentages of the joint mutual information. \label{Tab1}  }
 \begin{tabular}{cccccc} \toprule
 Condition & $I(Y ; B)$ &$ I(Y ; A)$ & $I(Y; B |A)$ & $I(Y ; A | B)$ & $II(Y ; B ; A)$ \\ \midrule
Control& 61.2 & 16.3 &  83.7 & 38.8 &26.9 \\
  & (52.1, 69.8) & (11.8, 19.5) & (80.5, 88.2) & (30.2, 47.9) &  (14.5, 30.0) \\ \hline
Baclofen& 77.6&  5.2 & 94.8 & 22.4  & 13.6 \\
 & (71.6, 86.3) & (4.3, 8.2) & (91.8, 95.7) & (13.7, 28.4) &  (9.6, 22.3) \\
 \bottomrule
 \end{tabular}
 \end{table}

From Table~\ref{Tab1} we find that $I(Y;B)$ increases on average when baclofen is present and that the sample distribution of values has shifted upwards while having approximately the same interquartile range. It is also noticeable that $I(Y;A)$ has decreased on average in the presence of baclofen and that the sample distribution of values has shifted downwards  with an approximate 50\% reduction in interquartile range. By considering $\eqref{norm}$, we find corresponding  changes  in the conditional mutual informations $I(Y;B|A)$ and $I(Y;A|B)$, which indicate, on average,  an increase in the conditional dependence between the AP output and the basal input along with a decrease in the conditional dependence between the AP output and the apical input.
These changes, which are associated with the presence of baclofen, show that, as expected, when there is inhibitory input to the distal apical dendrite, the AP output becomes more strongly related to the basal input and less strongly related to the apical input. It remains to be seen just how these changes are reflected in the differences between the components of the PIDs.

All of the thirty values of the interaction information are positive. Thus we can deduce the presence of at least some synergy {\it a priori} for all fifteen neurons under each condition for  PIDs that are guaranteed to possess nonnegative components.

\subsubsection{Comparison of PID components}

\begin{figure}[H]
\centering
\begin{tabular}{c}

\begin{subfigure}{\textwidth}\centering{\includegraphics[width =13 cm]{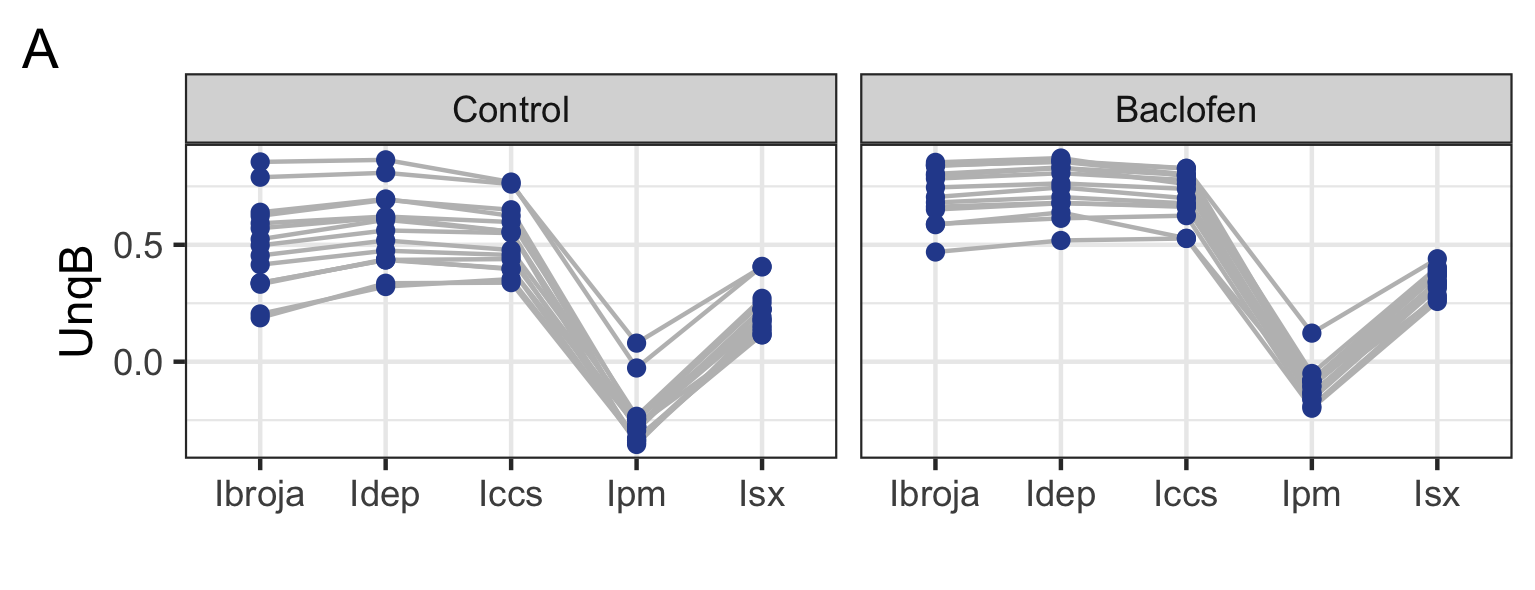}}  \label { } \end{subfigure}\\
\begin{subfigure}{\textwidth}\centering{\includegraphics[width = 13 cm]{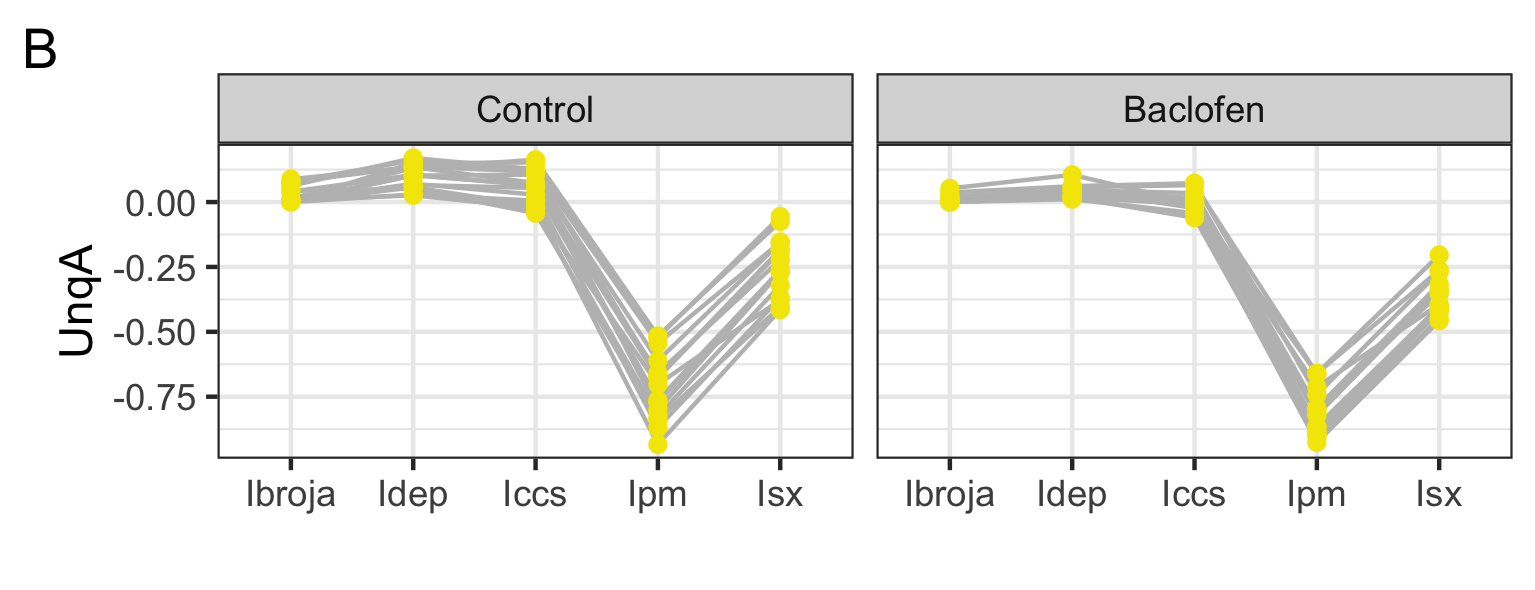}}  \label { } \end{subfigure}\\
\begin{subfigure}{\textwidth}\centering{\includegraphics[width = 13 cm]{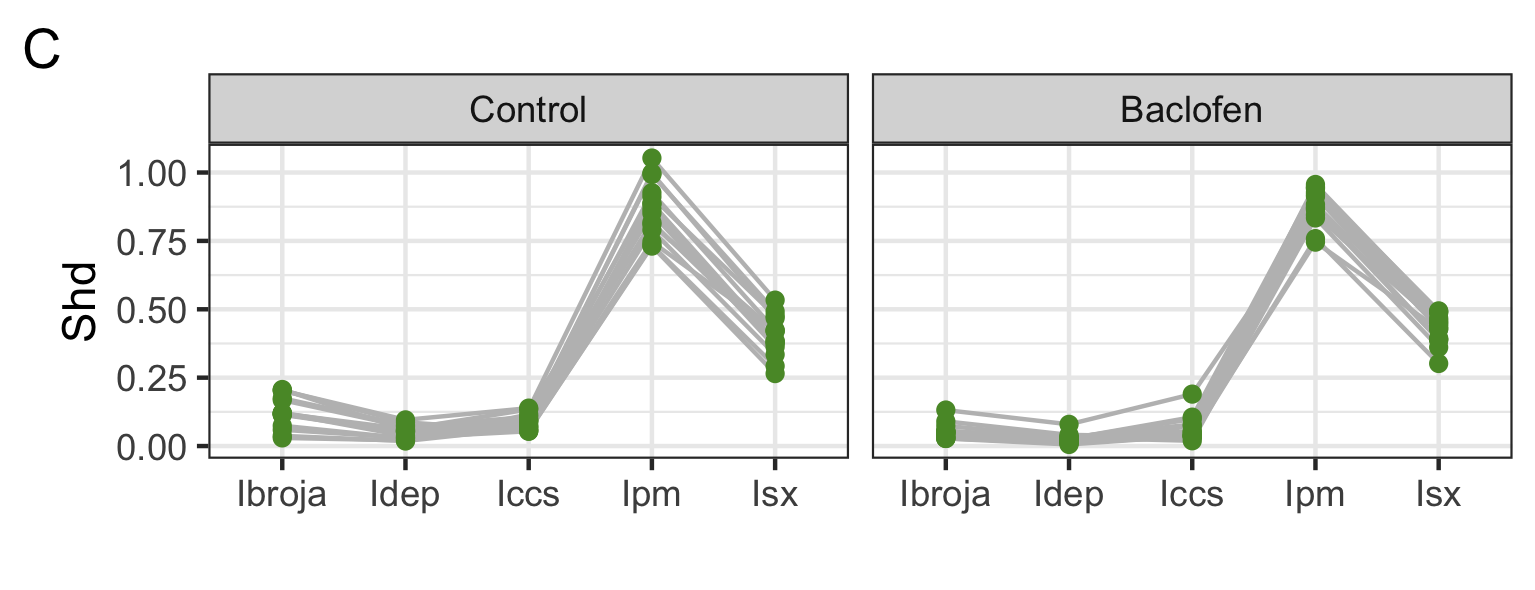}} \label { } \end{subfigure}\\
\begin{subfigure}{\textwidth}\centering{\includegraphics[width = 13 cm]{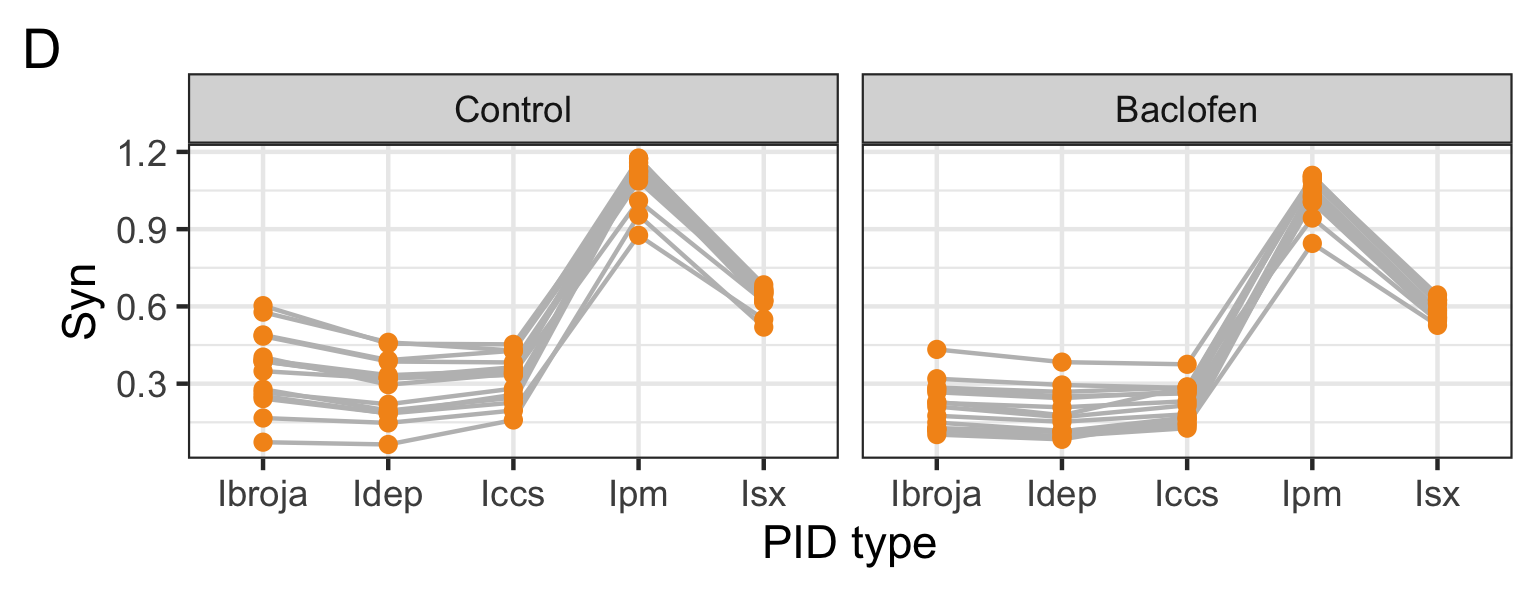}} \label { } \end{subfigure}

\end{tabular}
\caption{Physiological L5b neuronal recording data. Plots of  each PID component, connected by each neuron, for the five PID methods: A (UnqB), B (UnqA), C (Shd) and (D) Syn. For each neuron under each experimental condition, the values of the PID components are given as proportions of the respective joint mutual informations.}
  \label{fig3}
\end{figure}

The components of the five PIDs are plotted in Figure~\ref{fig3}. For each PID component, the values given by the Ibroja, Idep and Iccs appear to be reasonably similar for each of the Control and Baclofen conditions, although the Iccs method provides a small negative value of unique information due to the apical input for a few neurons. 
By way of contrast, the Ipm and Isx provide different ranges of values for each PID component, particularly evident in the plots of shared information and synergy where their ranges of values do not overlap at all with those of the PIDs Ibroja, Idep and Iccs, and they give much larger values for these components. In particular, both of the methods Ipm and Isx give very negative values of the unique information due to the apical input, much more strongly negative in the case of Ipm. Ipm also gives negative values in most cases  for the unique basal information; for Isx this happens only in one case. Also, the ranges of values provided by Ipm and Isx do not overlap at all except for the case of the unique information due to the basal input. The Ipm method also has values for synergy that are greater than 1, which suggests somewhat counterintuitively that more information is transmitted in the form of synergy than is available in the joint mutual information.

Summary statistics for the sample of fifteen neurons are given in Table 2. For each of the PID components, the median values for the Ipm and Isx PIDs are very different from those produced by the Ibroja, Idep and Iccs methods, being much smaller (or negative) for UnqB and UnqA and also much larger for the components Shd and Syn. On average, The Idep and Iccs PIDs generally have slightly larger values of the unique informations than does the Ibroja PID, and correspondingly lower values for Shd and Syn. It is clear from Table 2 and Figure~\ref{fig3} that, for these data sets, the Ipm and Isx methods produce remarkably different PIDs. If one were interested in estimating the actual value of each component then researchers using different methods would obtain very different values, including much larger estimates of shared information and of synergy! 

\setlength{\tabcolsep}{4pt}
\begin{table}[H]
\centering
\caption{Physiological L5b neuronal recording data. Summary statistics for 15 neurons.  PID components  are shown for each PID method and each experimental condition. The sample median (Md) and the sample quartiles (q$_{\rm L}$, q$_{\rm U}$) are stated as percentages of the joint mutual information.}
\begin{tabular}{lcrrrrrrrrrrr} \toprule
& & \multicolumn{5}{c}{Control} && \multicolumn{5}{c}{Baclofen} \\ \toprule
& & Ibroja &                 Idep & Iccs & Ipm & Isx &&           Ibroja & Idep & Iccs & Ipm & Isx  \\ \midrule \\
& q$_{\rm L}$ & 33.7 & 43.5 & 39.7 & $-33.1$ & 13.9 && 65.7& 65.8 & 64.3 & $-16.0$ & 13.6 \\
UnqB & Md & 49.7 &    51.9 & 47.7 & $-27.8 $& 18.5 && 74.5 & 74.6 & 70.0 &$ -12.7$& 19.2 \\
&  q$_{\rm U}$ &60.8 & 65.7& 61.1 & $ -24.7$& 28.3 && 82.0 & 84.3 & 78.4& $-8.2$ &  24.0 \\ \midrule \\

& q$_{\rm L}$ & 0.2 & 6.1 & 1.7 & $-79.8$ & $-33.3$ && 0.0& 1.9&$ -0.7$&$ -88.5$ & $-34.5$ \\
UnqA & Md & 0.7 & 10.3 & 6.7 & $-67.8$ & $-24.2$&& 0.7 & 4.0 & $-0.1$ &$ -81.7$ & $-22.2$ \\
&  q$_{\rm U}$ &5.3 & 13.6& 12.2 &  $-59.2$&$ -15.6$&& 1.9 &  5.5 & 2.1& $-72.7$ & $ -15.5$ \\ \midrule \\

& q$_{\rm L}$ & 19.4 & 3.9 & 6.3 & 80.0& 36.7 && 3.5& 1.8 & 3.6 & 85.5 & 38.8 \\
Shd & Md & 11.9 & 5.4 & 7.4 & 86.6& 38.3 && 5.0& 2.7 & 4.7 & 88.4& 42.9 \\
&  q$_{\rm U}$ & 17.1 & 6.6& 10.1 &  91.9& 43.5 && 5.5 & 3.4& 7.6& 94.2 &  46.0 \\ \midrule \\

& q$_{\rm L}$ & 25.9 & 20.3 & 26.9 & 109.3 & 62.3 && 12.7& 11.3& 16.7 & 101.4 & 55.8 \\
Syn & Md & 38.6 & 32.2 & 34.3 & 112.4 & 65.1 && 21.1 & 17.7 & 23.2 & 106.3 & 58.7\\
&  q$_{\rm U}$ &44.4 & 38.6& 39.2&  116.6& 66.2 && 27.2 & 25.9 & 28.5& 110.0 &  62.3 \\ 

\bottomrule
\end{tabular}
\end{table}

If the interest in the components were relative, however, and involved comparing PID components under different conditions then perhaps the dramatic differences between the Ipm and Isx methods and the other methods would  be somewhat attenuated, thus rendering the results produced by the different PID methods to be fairly similar in a relative sense, if not at an absolute level.

The purpose of the the study by Schulz et al.~\cite{SKBL} was to examine the effect of the GABA$_{\rm B}$ receptor-mediated dendritic inhibition on dendritic integration, and in particular whether it was associated with a change in synergy. This involves the examination of within-neuron differences of the synergy component, and so it means the comparison of the relative values of synergy in the absence and presence of local  baclofen application that activated GABA$_{\rm B}$ receptors in the distal apical dendrite. We now turn attention to these comparisons.

\subsubsection{Analysis of within-neuron differences in PID components}

\begin{figure}[H]
\centering
\includegraphics[width = 13 cm]{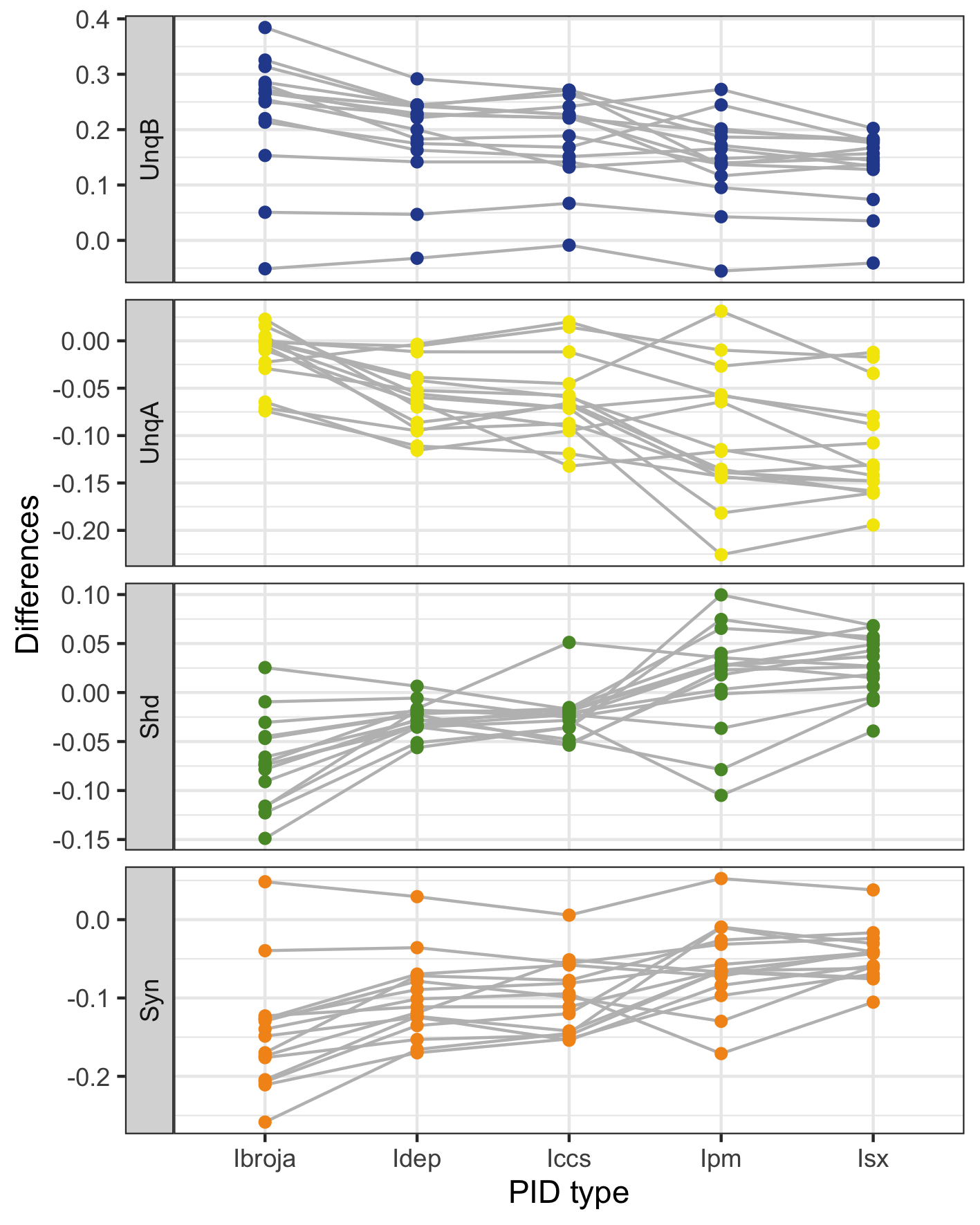}

\caption{Physiological L5b neuronal recording data. Within-neuron differences in each PID component for 15 neurons, taken as Baclofen minus Control. Different vertical scales are employed for each component.}
  \label{fig4}
\end{figure}

In Figure~\ref{fig4} the within-neuron differences in each PID component are plotted for each neuron.

 It is clear that the five PID methods produce differences which lie on the same scale. For UnqB, we see that the differences for all but one neuron are positive for all of the five PID methods, suggesting a general increase in UnqB in the presence of baclofen. For UnqA, most of the differences are negative for all five PIDs, thus indicating a general decrease in UnqA in the presence of baclofen.  The Shd component differences reveal a possible divergence of the PID methods: while the Ibroja, Idep and Iccs differences are almost all negative, those for Ipm  and Isx are mostly positive, thus suggesting a general increase rather than a general decrease. Apart from one neuron, the synergy differences are all negative, suggesting a general decrease in synergy in the presence of baclofen.

These observations on the plots of the component differences are also reflected in the summary statistics provided in Table 3.

\begin{table}[H]
\centering
\caption{Physiological L5b neuronal recording data.  Summary statistics of within-neuron differences for 15 neurons. The sample median (Md) and the sample quartiles (q$_{\rm L}$, q$_{\rm U}$) of the differences (taken as Baclofen minus Control) are provided for each PID component.  }
\begin{tabular}{llrrrrr} \toprule
& & Ibroja & Idep & Iccs& Ipm & Isx \\ \midrule

& q$_{\rm L}$ & 21.7& 16.9 & 14.6& 12.6 & 13.1  \\
UnqB & Md & 26.6 & 22.1 & 22.1 & 14.8 & 14.8 \\
& q$_{\rm U}$ &     24.2 &      24.2 &     23.4 &   19.2 &     17.8       \\ \midrule

& q$_{\rm L}$ & $~~~~-0.0$ & $ -9.0 $& $-8.9$ & $-14.1$ & $-15.3$  \\
UnqA & Md & $-0.0$ & $-6.2$ & $-6.8$ & $-11.6$ & $-13.5$ \\
& q$_{\rm U}$ &        0.0 &    $-4.0$ &     $ -5.1 $&   $ -5.7$ &  $ -8.4 $       \\  \midrule

& q$_{\rm L}$ & $-10.3$ & $-3.4$ & $-3.6$ & $0.1$ &  1.1  \\
Shd & Md & $-7.4$ & $-2.5$ & $-2.4$ & 2.5 &  2.7 \\
& q$_{\rm U}$ &       $ -4.6$    & $-1.8$ &    $-1.8 $ &  3.8 &        5.1   \\  \midrule

& q$_{\rm L}$ &$ -19.0$ & $-13.0 $&$ -14.4$ &$ -8.0 $&$ -6.6$ \\
Syn& Md &$ -14.9$ &$ -10.6$ & $-9.9$ & $-6.5$ & $-4.3$ \\
& q$_{\rm U}$ &     $ -12.6$&    $  -6.8$ &     $ -5.7$ &  $ -2.9$ &   $-3.6$       \\ \bottomrule
\end{tabular}
\end{table}

\subsubsection{Statistical significance?}
For the physiological L5b neuronal recording data~\cite{SKBL}, there is interest mainly in the synergy components. Suppose that five different researchers were to each use a different PID method and then apply a statistical test of the null hypothesis that the median value of synergy is the same in the absence and in the presence of baclofen. It turns out that all five researchers would find a significant reduction, on average, in  the synergy component (all p values are less than 0.001) when baclofen is present. Therefore, despite the dramatic differences between the PID results, both in the absence and in the presence of baclofen, all five researchers would arrive at the same formal statistical conclusion. 

Suppose, however, that interest lay  in the shared information component. For this component all five PIDs do not produce the same formal statistical conclusion. The researchers using Ibroja (P < 0.001), Idep (P < 0.001) and Iccs (P < 0.006) would all find a statistically significant reduction, on average, in the shared information in the presence of baclofen. On the other hand, the researcher using Ipm would declare that there is no statistically significant difference, on average, in this component when baclofen is introduced (P = 0.2), and with Isx the researcher would declare a statistically significant {\it increase}, on average, in the shared component (P < 0.006).

\subsubsection{Unique Information Asymmetry}
\begin{figure}[H]
\centering
\includegraphics[width = 5in]{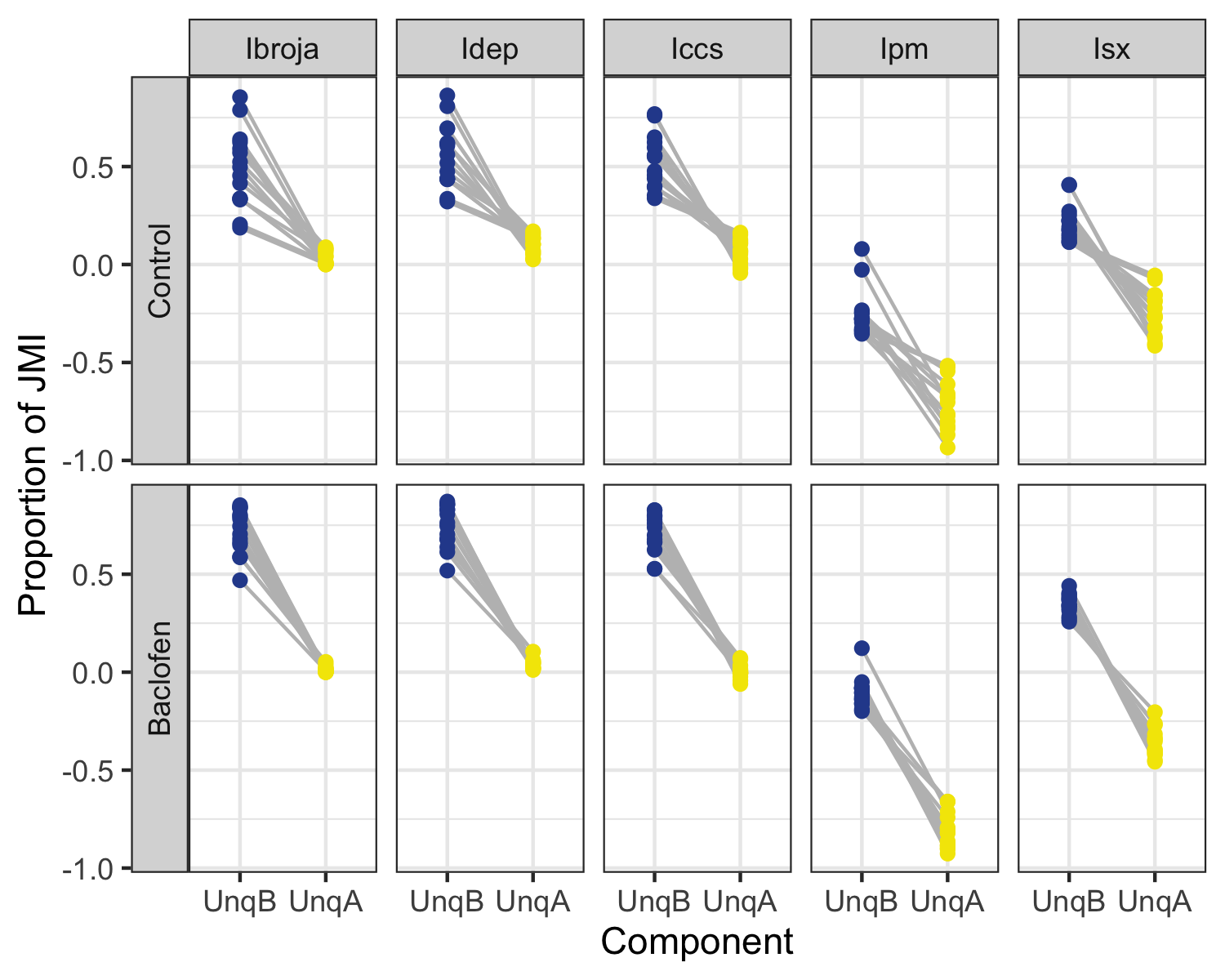}
\caption{Physiological L5b neuronal recording data.  A plot of the unique information asymmetries  provided by each of the five PIDs for each experimental condition for 15 neurons. }
\label{fig5}
\end{figure}

Recall from Section 2.3 that the unique information asymmetry (UIA) is defined as UnqB - UnqA. For a given probability distribution the UIA has the same value for every PID. Figure~\ref{fig5} shows the the UnqB and UnqA values for each experimental condition and each PID. The 15 neurons have positive values of the UIA under each experimental condition. Despite the fact that the Iccs PID has a few very small negative values for UnqA it appears that the three PIDs, Ibroja, Idep and Iccs, have very similar patterns to each other under each of the experimental conditions. On the other hand Ipm and Isx express the asymmetries very differently, and even differently from each other. Apart from one of the 15 neurons, Ipm has negative values for UnqB and more negative values for UnqA. Hence the asymmetries are being expressed in terms of there being much more unique apical misinformation than unique basal misinformation. For each of the 15 neurons, Isx has positive values for UnqB and negative values for UnqA. Therefore, it expresses the asymmetry as  a balance of unique basal information and unique apical misinformation, with the former being larger for some neurons and smaller for others.

One property of apical amplification (CSS3) in a neuron is that there is no or little unique apical information or misinformation  in a PID, coupled with the requirement that  synergy or mechanistic shared information, or both, are present. It is clear in Figure~\ref{fig5} that both Ipm and Isx mostly have large unique misinformation components and so, while they do have large values of synergy, their PIDs are generally not compatible with property CCS3. Therefore, we focus on the other three PIDs. For Ibroja, most of the neurons have small nonnegative unique apical components as well as appreciable synergy components (see Figure~\ref{fig3}) and so they provide some evidence of apical amplification. The Idep and Iccs PIDs tend to produce larger values for the unique informations than those obtained using Ibroja, but in the Control condition a few neurons have small unique apical components, and this is more markedly the case in the presence of baclofen. From Figure~\ref{fig3}, we see that several of these neurons also have appreciable values for synergy. Thus, some evidence of apical amplification is given by these neurons when using the Idep and Iccs methods but the Ibroja method provides the strongest support for apical amplification.

Some summary statistics of the UIA values and their differences, with Bonferroni-corrected p values,  are provided in Table~\ref{Tab4}. The UIA is significantly positive, on average, under the control condition, and also in the presence of baclofen. When baclofen is present, the UIA is significantly larger, on average, than in the control condition. For the Ibroja, Idep and Iccs PIDs, these results taken together with  Figure~\ref{fig4}, which shows for 14 of the neurons that in the presence of baclofen there is an increase in the transmission of unique basal information,  coupled with a decrease in shared information and synergy,    confirm the finding in~\cite{SKBL} that `GABA$_{\rm B}$R-mediated inhibition shifts the balance towards somatic control of AP output and potently decreases apical amplification'.

\setlength{\tabcolsep}{6pt}
\textcolor{blue}{\begin{table}[H]
\centering
\caption{Physiological L5b neuronal recording data.  Summary statistics of the values of unique information asymmetry for 15 neurons. The sample median (Md) and the sample quartiles (q$_{\rm L}$, q$_{\rm U}$) are stated as percentages of the joint mutual information. The differences between the unique information asymmetries are taken as Baclofen minus Control. The quoted P values have been Bonferroni-corrected since three simultaneous tests have been performed. } \label{Tab4}
\begin{tabular}{cccc} \toprule
 & Control & Baclofen & Difference \\ \midrule
 q$_{\rm L}$ &  33.3 & 64.7  &   21.8   \\
 Md &  44.6&  74.5& 27.9 \\
 q$_{\rm U}$ & 60.7& 80.1& 33.4 \\ \midrule
 P & < 0.0002 & <0.0002  & < 0.0004 \\
\bottomrule
\end{tabular}
\end{table} }

\subsection{Simulated data from a detailed compartmental model}
Shai et al.~\cite{Shai} reported simulations of a L5b model neuron that was based on a model originally fitted to data recorded from the rat somatosensory cortex by Hay et al.~\cite{HHSMS} and then adapted to recordings from the adult mouse visual cortex by manual manipulation of dendritic calcium and I$_{\rm H}$ conductance parameters.
NMDA/AMPA synapses were randomly distributed across the tuft and basal dendrites ranging in number from 0 to 300, in steps of 10,  in the basal dendrites and 0 to 200, in steps of 10,  in the apical dendrites. While many bursts of APs were observed,  information regarding their occurrence was not recorded and so is unavailable in the data set (Adam Shai, {\it personal communication}). Hence we work with spike counts.

\begin{figure}[H]
\centering
\includegraphics[width =4in]{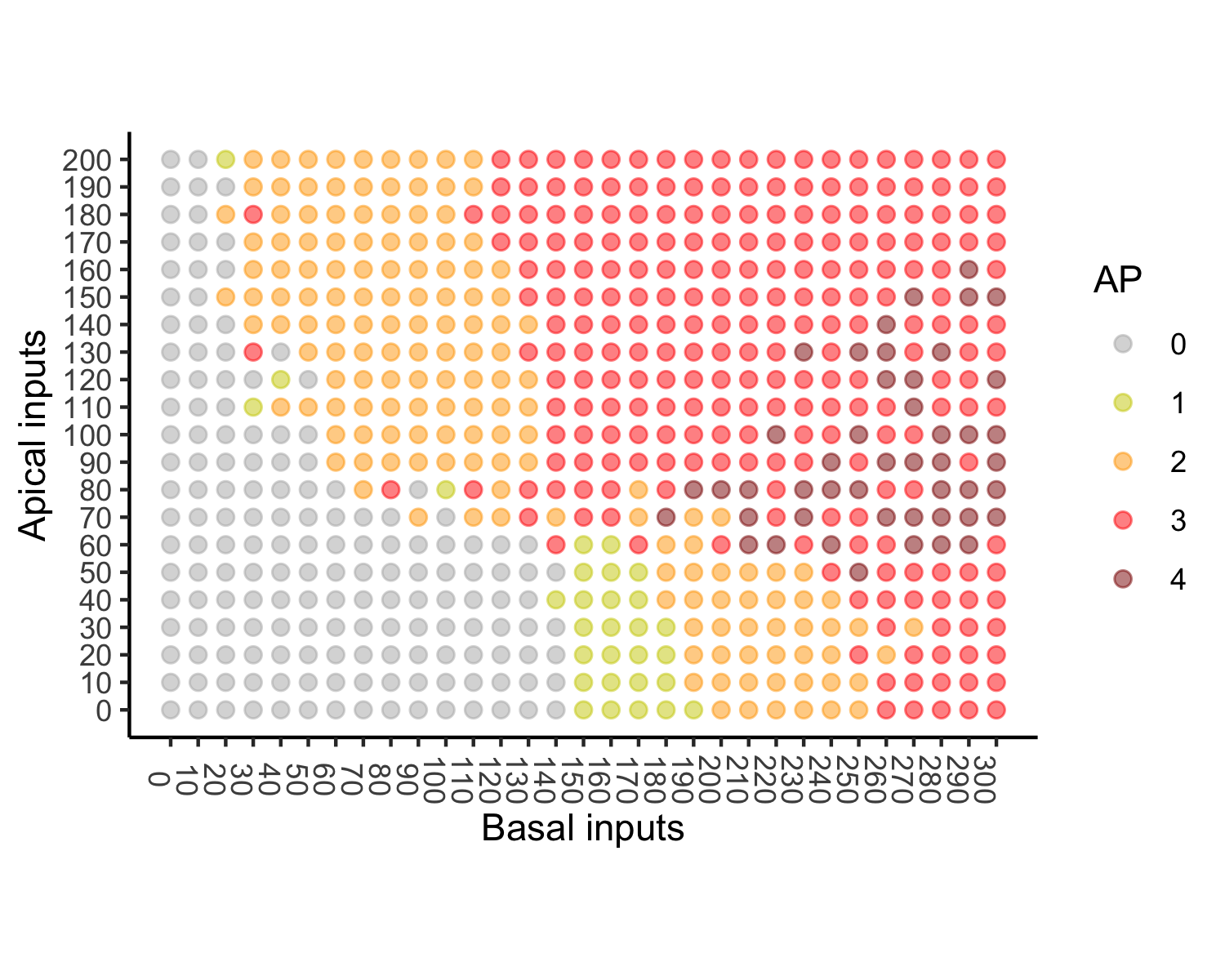}

\caption{Simulated mouse L5b neuron model data. The number of action potentials emitted are provided for 651 combinations of the numbers of basal and apical inputs to the cell. The number of basal inputs ranges from 0 to 300 in steps of 10. The number of apical inputs ranges from 0 to 200 in steps of 10.}
\label{spikes}
\end{figure}
Information regarding the observed numbers of APs for the combinations of numbers of basal and apical inputs used in the study is provided in Figure~\ref{spikes}. 
No APs are evoked by apical inputs when there are 0 or 10 basal inputs, but when there are no apical inputs APs occur provided that the number of basal inputs is at least 160. This suggests that the basal input is driving when the number of apical inputs is very low. When the number of basal inputs is very low (30-50) APs occur as long as the number of apical inputs is 110 or larger, suggesting that apical inputs may be the more effective driver of AP output under certain circumstances.  When the numbers of basal and apical inputs are both large then at least 3 APs occur.

Since there are relatively few combination with 1 or 4 APs (see Figure~\ref{spikes}) the categories of the output variable $Y$ were taken to be 0, 1-2, 3-4 APs. The numbers of basal and apical inputs were not categorised. Each of the 651 observed combinations was given a probability of $1/651$, with the remaining possibilities having a probability of 0, thus creating a 31 values for the basal input, $B$, 21 values for the apical input, $A$, and 3 possible values for AP count, $Y$, in a 31 by 21 by 3 probability distribution. 

Several classical information measures were computed for this probability distribution.
\begin{table}[H]
\centering
\caption{Simulated mouse L5b neuron model data. Estimated mutual  information measures.
Some estimated classical mutual information measures (the unit is bit), given to two decimal places. The numbers in parentheses are the values of the measures as a percentage of the joint mutual information. \label{Tab5}  }
\centering
\begin{tabular}{ccccccc}
\toprule
$I(Y; B)$ & $I(Y;A)$ & $I(Y;B|A)$ & $I(Y;A|B)$ & $I(Y; B, A)$ & $II(Y; B; A)$ & $H(Y)$\\ \midrule
0.74 &0.16 & 1.37 & 0.79& 1.54 & 0.63& 1.54\\
(48.3) & (10.7) & (89.3) & (51.7) & (100) & (41.0) & (100)\\

\bottomrule
\end{tabular}
\end{table}

In this 31 by 21 by 3 system, the joint mutual information $I(Y; B, A)$ is  1.54 bits, while the difference  $I(Y; B) - I(Y; A)$ is 0.58 bit. Therefore  in any  PID the unique information due to basal input will be larger than that contributed by apical input   by 0.58 bit.  The interaction information in this system is 0.63 bit, which is 41\% of the joint mutual information. Thus without performing a PID we can deduce, for any PID having nonnegative components, that  at least 41\% of the mutual information between the output $Y$ and the inputs $(B, A)$  will be due to synergy. 

\begin{figure}[H]
\centering
 \includegraphics[width = 3in, height =0.5in]{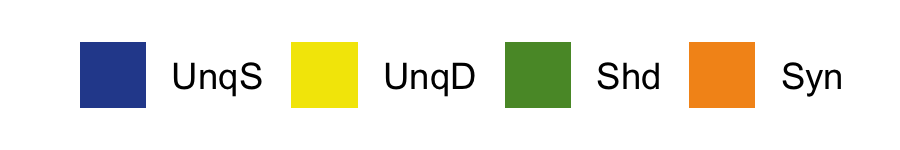}

 \includegraphics[width =4in]{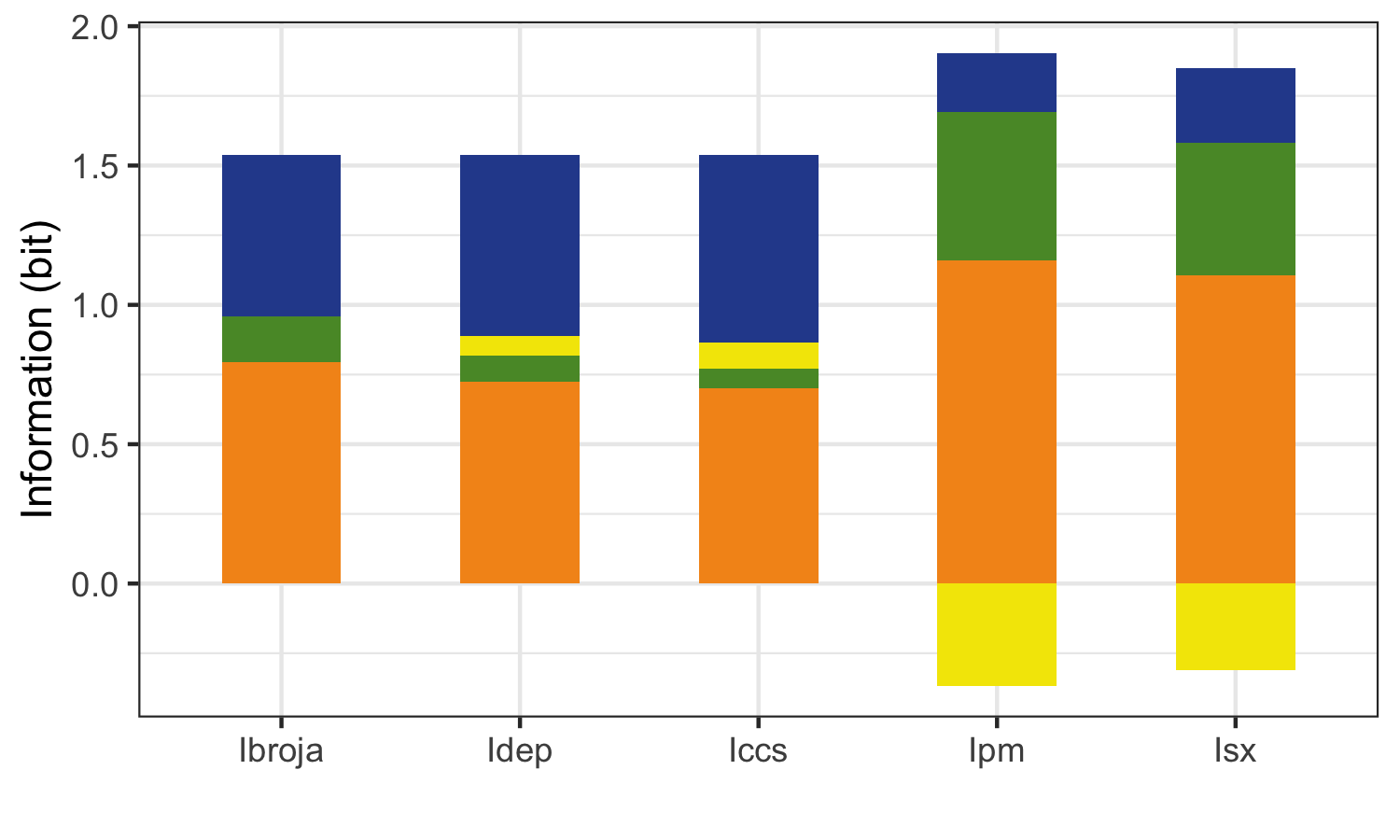}

\caption{Simulated mouse L5b neuron model data. Stacked bar plots showing the values of the components for each of the five PIDs using the full data set }
\label{ShaiBar}
\end{figure}

To obtain the actual values of the partial information components, the five PIDs were applied to the whole data set and the results are given in  Figure~\ref{ShaiBar}. 

  We now describe the bar plots in terms of percentages of the joint mutual information. As expected all five PIDs reveal the asymmetry between the unique basal and apical components, possibly due to the disparity in numbers of apical and basal inputs -- but see Figures~\ref{fig8}, \ref{fig9}  which contain a combination where both basal and apical have the same range of inputs: 0-200. There  are differences in how this asymmetry is expressed. The Ibroja PID has 37.7\% of the joint mutual information transmitted as information unique to the basal input and no unique information due to the apical input. For the Idep PID, the respective numbers are 42.2\% and 4.5\%, while for the Iccs PID they are 43.7\% and 6.1\%. This suggests for all three PIDs that the basal input is primarily driving while the apical input is mostly amplifying. Thus these three PIDs express the asymmetry in a similar manner, with Ibroja providing the strongest suggestion of apical amplification.

On the other hand the PIDs Ipm and Isx express the asymmetry rather differently. For Ipm, 13.8\% is transmitted as information unique to the basal input, whereas 23.9\% is transmitted as unique apical misinformation. The corresponding numbers for Isx are 17.4\% and 20.3\%. These two PIDs express the asymmetry in a similar manner. The numbers suggest that both the basal and apical inputs are  driving, with the basal input transmitting information while the apical input is contributing a larger percentage as misinformation.  For all five PIDs a large percentage of the joint mutual information is transmitted as synergy, with much larger percentages for the PIDs Ipm and Isx than for the other three PIDs. Also, the percentage of information transmitted as shared information is much larger for the PIDs Ipm and Isx.

\subsubsection{PID analysis for varying strengths of basal and apical input}
In previous work on cooperative context-sensitivity~\cite{KIDP, KP2020}, by utilising pre-defined probability models and particular transfer functions, it was possible to explore ideal properties, as defined in the Methods section. In order to investigate such matters here with realistic data, we consider increasing subsets of numbers of basal and apical inputs, from 0-100 to 0-200 for each of the basal and apical inputs. We think of a range which has larger numbers of inputs as being stronger than a range with a smaller number of inputs; so the range 0-130 is viewed as being stronger input that the range 0-100, and if the ranges of basal and apical inputs are both, say, 0-150 we consider the strengths of the basal and apical inputs to be equal.

\begin{figure}[H]
 \begin{center}
 \includegraphics[width = 3in, height =0.5in]{Nbroja_leg.png}
 \end{center}
\centering
\begin{tabular}{p{2cm}p{4cm}p{4cm}p{3cm}p{2.5cm}}
&Ibroja & Idep & Iccs&\\
\end{tabular}
\includegraphics[width =13.5 cm]{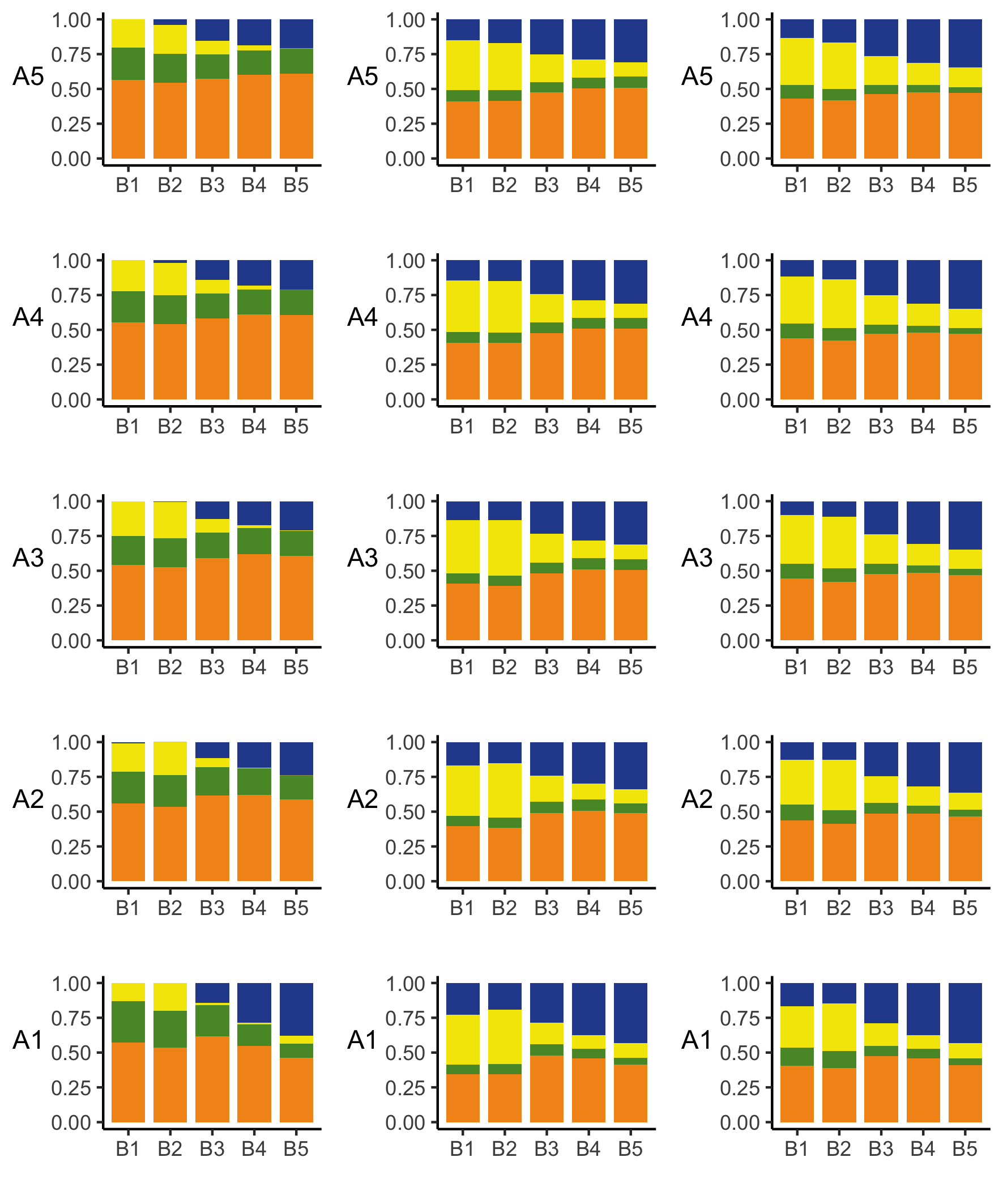}
\caption{Simulated mouse L5b neuron model data. Ibroja (left column), Idep (middle column) and Iccs (right column) PIDs for various combinations of increasing ranges of basal inputs and increasing ranges of apical inputs: B1(0-100), B2(0-130), B3(0-150), B4(0-170), B5(0-200) and A1(0-100), A2(0-130), A3(0-150), A4(0-170), A5(0-200).}
\label{fig8}
\end{figure}

\begin{figure}[H]
 \begin{center}
 \includegraphics[width = 3in, height =0.5in]{Nbroja_leg.png}
 \end{center}
\centering
\begin{tabular}{p{4.5cm}p{4cm}p{3cm}p{3cm}}
 & Ipm & Isx&\\
\end{tabular}
\includegraphics[width =9 cm]{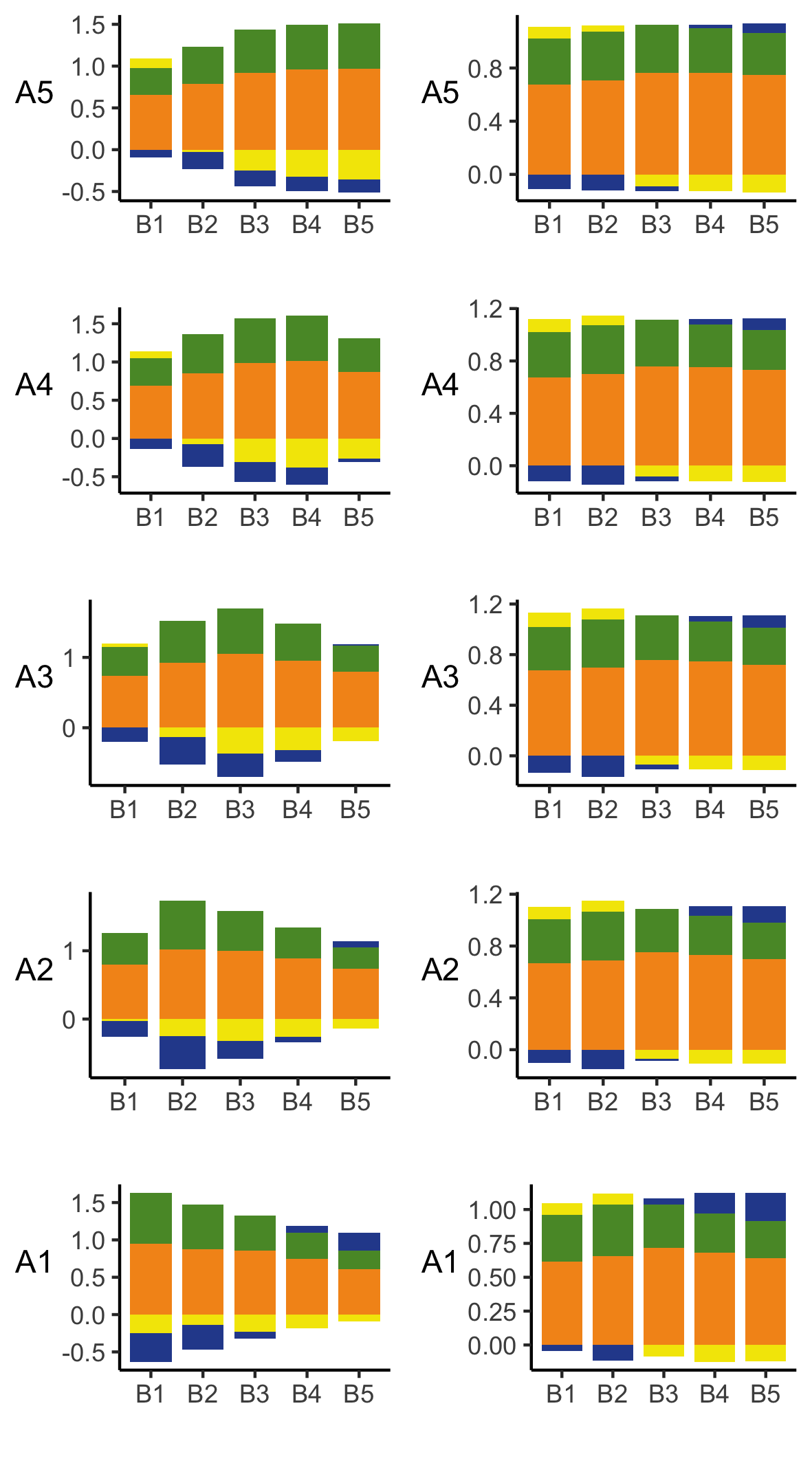}
\caption{Simulated mouse L5b neuron model data. Ipm (left column) and Isx (right column) PIDs for various combinations of increasing ranges of basal inputs and increasing ranges of apical inputs: B1(0-100), B2(0-130), B3(0-150), B4(0-170), B5(0-200) and A1(0-100), A2(0-130), A3(0-150), A4(0-170), A5(0-200). }
\label{fig9}
\end{figure}

We take the large range 0-100 as a baseline as there is no information in several smaller ranges since there are no APs (Figure~\ref{spikes}). Starting with the range 0-100 an additional 10 units were added incrementally until the range 0-200 was reached. 
Therefore, when basal and apical both have the range 0 -100, we see from Figure~\ref{spikes} that  there are 11 distinct basal inputs and 11 distinct apical inputs, and for each of the 121 combinations there are three possible values for the output.  Therefore the PIDs are based on an 11 by 11 by 3 probability distribution, with each observed combination having an equal probability of 1/121, with the remaining combinations having probability zero. Similarly, when the apical range is 0-100 and the basal range is 0-200 the PIDs are based on a 21 by 11 by 3 probability distribution, with each observed combination having equal probability 1/231, with the rest having probability zero. When the ranges are both 0-200 the PIDs are based on a 21 by 21 by 3 probability distribution with each observed cell having probability 1/441, with the rest having probability zero. Thus there are 121 different combinations of ranges of basal and apical input. Given the different sizes of the probability distributions and the resulting differences in the values of the joint mutual information, the components of a PID in each combination were normalised by dividing by the joint mutual information for that combination. A representative sample of the 121 PIDs for each of the five methods is displayed in Figures~\ref{fig8}, \ref{fig9}.

Focusing on the Ibroja results in Figure~\ref{fig8}, we notice that there is a large synergy component in each combination, as well as an appreciable level of shared information. These levels of synergy and shared information appear to be fairly constant in all combinations for which the apical range is 0-130 or greater. 

When the apical input is 0-100, however, we do notice some changes in the shared information and the synergy. As the basal range increases there is an increase in both of these components until the range 0-150 and thereafter a small decrease in both as the basal range increases.

We now comment in the changes in the unique informations and their relative values for these PIDs.  When the basal input is 0-130, or lower, this asymmetry is negative for these subsets and there is very little unique basal information at all. On the other hand, when the basal input is 0-170 or greater the asymmetry is positive and there is very little unique apical information. 
The negative  asymmetry is present because the number of basal inputs is not sufficient to drive APs, whereas apical inputs are more effective at driving APs. With regard to the the positive asymmetry, now the situation changes, because basal inputs do drive APs; however, they do this in a more graded fashion than apical inputs. 
See Figure~\ref{spikes}.

When the basal input is 0-150 we find that the asymmetry becomes positive, and one could say that by considering all the basal input ranges there is a unique information asymmetry bifurcation which happens when the number of basal inputs increases from 140 to 150, irrespective of the strength of the apical input.  Thus these results reveal a much more diverse picture than the PID analysis of the whole dataset. For basal input up to 0-140 we say that the apical input is driving, whereas this begins to reverse at 0-150 and more strongly so for the larger basal input ranges. It is interesting that these patterns of results are not obtained if one were to reverse the roles of basal and apical and consider increasing apical strength; this reveals a fundamental asymmetry within the distributions.

A unique information asymmetry can also be seen when the ranges of basal and apical ranges are equal. For low values of both basal and apical there is apical drive and this changes to basal drive when the numbers of basal and apical inputs {\it both} change from 140 to 150. Even though the basal and apical strengths are equal, we find that the basal input comes to dominate in terms of unique information as the common strength increases.

These revelations also apply to the results obtained using the Idep and Iccs PIDs, although they both tend to produce larger values for the unique information components.

The Ipm and Isx results are given in Figure~\ref{fig9}. The comments regarding the unique information asymmetry bifurcation hold also for these PIDs due to the fact that unique information asymmetry is the same for all PIDs. It is expressed very differently, however. With Ipm, both unique informations are generally negative, so the asymmetry is described as the presence of  more unique basal misinformation switching to more unique apical misinformation. The Isx PID generally expresses the bifurcation in UIA as a mixture of unique apical information and unique basal misinformation changing to a mixture of unique basal information and unique apical misinformation.

\subsubsection{Cooperative context-sensitivity as revealed by PID analyses}
In these experiments, we add further basal ranges to the previous increasing ranges of basal inputs considered in Section 3.3, up to 0-300,  but now we consider three fixed apical ranges with a view to assessing the effect of different fixed strengths of apical input on the basal distributions of the PID components. From Figure~\ref{spikes}, we see that for apical ranges 110 -150 and 160-200, there are 5 distinct input values, and so the probability distributions range from 11 by 5 by 3 (for basal 0-100), with equal probability 1/55,  to 31 by 5 by 3 (for basal  0-300), with equal probability 1/155 for the observed combinations and zero for the rest. The PID components for each combination of input ranges are normalised by the joint mutual information for that combination.

\begin{figure}[H]
 \begin{center}
 \includegraphics[width = 3in, height =0.5in]{Nbroja_leg.png}
 \end{center}

\centering

\begin{tabular}{p{2cm}p{4cm}p{4cm}p{3cm}p{2.5cm}}
&Ibroja & Idep & Iccs&\\
\end{tabular}
\includegraphics[width =13.5 cm]{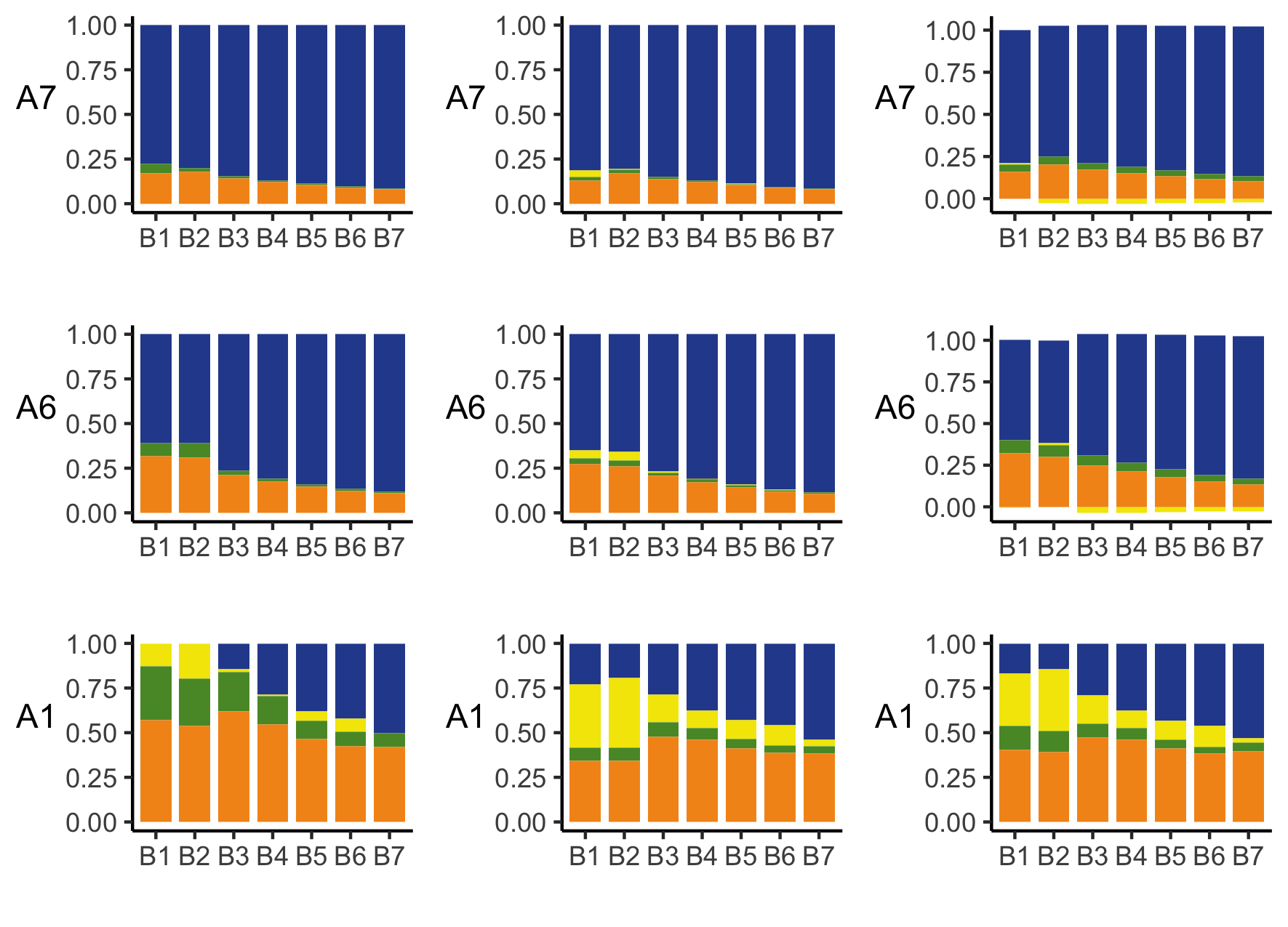}
\caption{Simulated mouse L5b neuron model data. Ibroja (left column), Idep (middle column) and Iccs (right column) PIDs for various combinations of increasing ranges of basal inputs and three fixed  ranges of apical inputs: B1(0-100), B2(0-130), B3(0-150), B4(0-170), B5(0-200), B6(0-250), B7(0-300) and A1(0-100), A6(110-150) and A7(160-200).  }
\label{fig10}
\end{figure}

The results obtained with the Ibroja, Idep and Iccs PIDs are given in Figure~\ref{fig10}, and those for the Ipm and Isx PIDs are in Figure~\ref{fig11}.
We now discuss the plots in Figs.~\ref{spikes}, \ref{fig10}, \ref{fig11} with regard to the ideal  properties of cooperative context-sensitivity defined in the Methods section, with the basal input as the `drive' and the apical input as the `context'. 
\subsubsection*{Properties CSS1 and CSS2}
In Figure~\ref{spikes} we find, in the absence of apical input, that APs are emitted when the number of basal inputs is at least 150. This shows that the basal input is sufficient for the output to transmit information about the input in the absence of context. Thus property CSS1 holds: $B$ is sufficient and $A$ is not necessary. When there is no basal input we see that no APs are emitted. Thus the apical input is not sufficient for information transmission and the basal input is necessary, and therefore property CSS2 holds. 

\subsubsection*{Property CSS3}
In Figure~\ref{fig10}, when the apical inputs are in ranges A6 and A7,  and the basal inputs are in ranges B1-B7,  the Ibroja, Idep and Iccs PIDs have large unique basal components as well as zero or small unique apical components, and synergy and some shared information are present; hence the these PIDs are consistent with property CSS3.  When the apical inputs range from 0-100, only the Ibroja PIDs for 0-150, 0-170 and 0-300 basal inputs are consistent with CSS3, while for the Idep and Iccs PIDs this is so only when the basal input range is 0-300.
In Figure~\ref{fig11}, the Ipm PID satisfies property CSS3 mainly when the basal input ranges are 0-200, 0-250 and 0-300, for all three ranges of apical input. The Isx PID does not have small components for unique apical information or misinformation and hence it does not produce results that are consistent with property CSS3. Therefore, property CSS3 holds most widely for the Ibroja PID,  less so for Idep and Iccs, for several basal-apical combinations with the Ipm PID and not at all for Isx. 

\subsubsection*{Property CSS4}
In each of the probability distributions considered which are defined in terms of combinations of ranges of basal and apical inputs, $B$ and $A$ are marginally independent, and so the source shared information, ShdS, is equal to zero. Therefore the shared information components describe mechanistic shared information. For Ibroja, in Figure~\ref{fig10}, when the apical input range is 0-100, and  the strength of the basal input increases from 0-130 to 0-150 we see that  the combined value of the UnqB, Shd and Syn information components increases, thus increasing the transmission of information about the basal input.

\begin{figure}[H]
 \begin{center}
 \includegraphics[width = 3in, height =0.5in]{Nbroja_leg.png}
 \end{center}
\centering
\begin{tabular}{p{4.5cm}p{4cm}p{3cm}p{3cm}}
 & Ipm & Isx&\\
\end{tabular}
\includegraphics[width =9 cm]{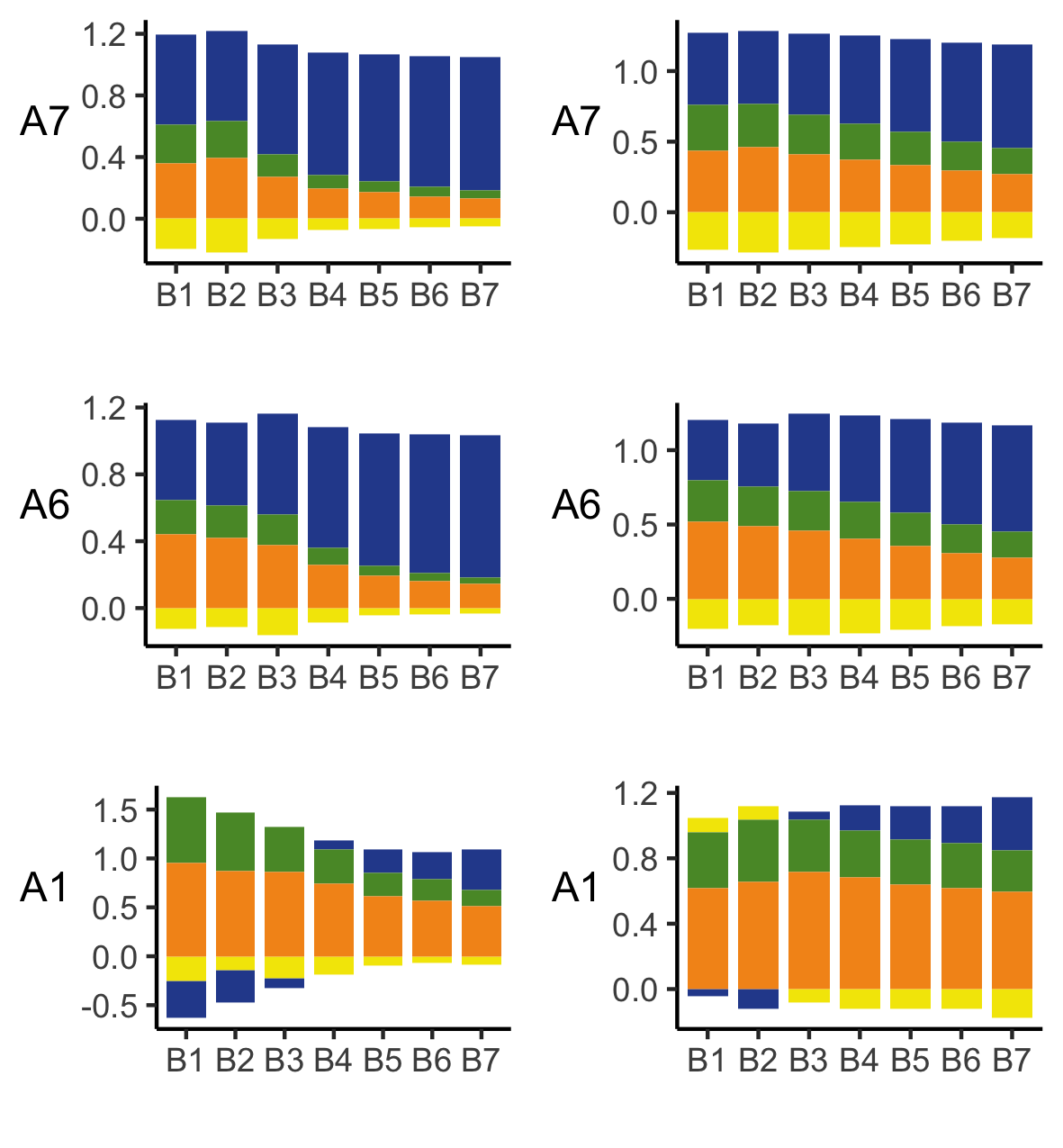}
\caption{Simulated mouse L5b neuron model data. Ipm (left column) and Isx (right column) PIDs for various combinations of increasing ranges of basal inputs and three fixed  ranges of apical inputs: B1(0-100), B2(0-130), B3(0-150), B4(0-170), B5(0-200), B6(0-250), B7(0-300) and A1(0-100), A6(110-150) and A7(160-200).  }
\label{fig11}
\end{figure}

 As the strength of the basal input is further increased we find that the shared and synergistic components generally decrease. This provides support for property CSS4. These observations hold also for the Idep PID components as well as the Iccs components. In Figure~\ref{fig11}, the Ipm PID does not show the increase in the combined value of the information components Shd and Syn (UnqB is misinformation) as the basal strength is increased from 0-130 to 0-150. As the strength  of the basal input is increased we do find, however, the same pattern of decreasing synergy and shared information as given by Ibroja, Idep and Iccs. Hence Ipm is partially consistent with property CCS4.  The Isx PIDs show the same characteristics that were shown with Ibroja, Idep and Iccs and so it is consistent with property CCS4.

\section{Conclusions}
\subsection*{Rat somatosensory cortical L5b pyramidal neuron recording data}
The PID analyses reveal that the Ibroja, Idep and Iccs methods produce broadly similar PIDs for the 15 neurons under each experimental condition, whereas the Ipm and Isx methods produce components which have very different values than those given by the Ibroja, Idep and Iccs methods. In particular, the Ipm and Isx methods produce much larger estimates of shared information and synergy. The Ipm method even produces some values  for synergy that are larger than the joint mutual information, which seems nonsensical! 

When the relative values of the PID components are considered - as with the within-neuron differences in the PID components used in the investigation of the effect of baclofen - then these differences can be considered on the same scale for all five PID methods, and the results are generally more similar, although not for the shared information component. Statistical testing shows that for the synergy component five independent researchers, each using one of the five methods, would arrive at the same formal statistical conclusion. Were they to consider the shared information, however, the researchers using the Ipm or Isx methods would reach a different formal statistical conclusion than obtained by those using  the Ibroja, Idep and Iccs methods.

While the values of the unique information asymmetry are the same for all five methods, the asymmetry is expressed in different ways. The Ibroja, Idep and Iccs methods all exhibit strong basal drive and there is evidence of apical amplification for several neurons. Examination of the within-neuron differences and statistical testing conducted on the asymmetry values provide support for the conclusion in ~\cite{SKBL} regarding the effect of GABA$_{\rm B}$R-mediated inhibition. Neither conclusion applies to the Ipm and Isx methods.

As to the question of which method(s) to rely on it seems wise, for probability distributions of the type considered in this study,  to employ the Ibroja, Idep and Iccs methods since they give broadly similar results, rather than the  Ipm or Isx method.

\subsection*{Simulated mouse L5b neuron model data}
The PID analyses of the full dataset again reveal differences among the five methods, with the Ibroja, Idep and Iccs decompositions again being broadly similar. The Ipm and Isx methods transmit higher percentages of the information as synergy and shared information and  an appreciable percentage as unique apical misinformation that is larger in magnitude  than the transmitted unique basal information.

A richer picture emerges when various subsets of the data are analysed. When the basal and apical inputs are treated on an equal footing, and various combinations of strengths of basal and apical inputs considered, we find that there is a bifurcation in unique information asymmetry for all PIDs. While the values of the asymmetries are fixed by classical measures of mutual information, the nature of the asymmetries is only revealed by the PIDs. For Ibroja, Idep and Iccs, we find that as the strength of the basal input increases, to the extent that it is sufficient to drive AP output,  there is a switch from apical drive to basal drive, and that this occurs at the same strength of basal input for every strength of apical input. We also find that this bifurcation happens when we consider combinations where the basal and apical strengths are equal. The Ipm and Isx PIDs express the asymmetry in terms of combinations of basal and apical misinformation or a combination of a unique information with a unique misinformation.
 
In a second exploration of subsets, increasing basal strengths were considered for three fixed apical strengths. With regards to cooperative context-sensitivity, we find that all five PIDs provide at least some support for the ideal properties. The Ibroja PID satisfies the properties to the fullest extent, with Idep and Iccs close behind. Ipm and Isx provide partial support.

Both investigated data sets have different intrinsic limitations. In the first study, direct current injection was used as an experimental approximation of synaptic inputs. The model neuron of the second study is expected to provide limited accuracy in the precise AP number evoked by synaptic inputs due to the intrinsic difficulties in modeling the fast underlying conductances appropriately~\cite{HHSMS}. However, our analyses converge at the conclusion that apical dendritic inputs may mainly contribute to synergy, i.e. have a modulatory role, rather than driving output information. The reason for this is that, in the investigated pyramidal neurons, apical dendritic inputs are bound to recruit an amplifying Ca$^{2+}$ spike mechanism in the apical dendrite if they were to activate somatic APs directly. Therefore, apical dendritic inputs cannot provide the graded impact on AP output that basal dendritic inputs do. We can conclude that under these circumstances, the role of apical dendritic inputs is largely restricted to amplifying output rather than driving output information.

\begin{flushleft}
{\large\bf Abbreviations}
\end{flushleft}

\begin{flushleft}
{Partial Information Decomposition}
\end{flushleft}

\begin{description}[labelsep =1cm, align =left, labelwidth = 1.2cm,  labelindent =0.2cm]

\item[PID] \parbox[t]{\length}{Partial Information Decomposition, with components UnqB, UnqA, Shd and Syn, defined in Section 2.3}\\
\item[Ibroja] \parbox[t]{\length}{The PID developed by Bertschinger et al.~\cite{BERT} }
\item[Idep] \parbox[t]{\length}{The PID developed by James et al.~\cite{James} }
\item[Iccs] \parbox[t]{\length}{The PID developed by Ince~\cite{RI} }
\item[Ipm] \parbox[t]{\length}{The PID developed by Finn \& Lizier.~\cite{FL} }
\item[Isx] \parbox[t]{\length}{The PID developed by  Makkeh et al.~\cite{MGW} }
\end{description}

\begin{flushleft}
{Others}
\end{flushleft}

\begin{description} [labelsep = 1cm, align =left, labelwidth =1.2cm,  labelindent = 0.2cm]
\item[AMPA]  \parbox[t]{\length}{ $\alpha$-amino-3-hydroxy-5-methyl-4-isoxazolepropionic acid }
\item[AP] \parbox[t]{\length} {Action potential  }
\item[GABA]  \parbox[t]{\length}{gamma-aminobutyric acid}
\item[GABA$_{\rm {\bf B}}$]  \parbox[t]{\length}{A GABA$_{\rm B}$ receptor is a G-protein receptor, or metabotropic receptor}

\item[JMI]  \parbox[t]{\length}{Joint mutual information between the output $Y$ and the inputs $(B, A)$, as defined in Section 2.2}
\item[L5b] \parbox[t]{\length}{Layer 5b }

\item[NMDA]  \parbox[t]{\length}{N-methyl-D-aspartate  }

\item[UIA]  \parbox[t]{\length}{Unique information asymmetry, as defined in Section 2.4}
\end{description}


\begin{thebibliography}{10}
\bibitem{WB}
Williams, P.L.; Beer, R.D.  Nonnegative decomposition of multivariate information. arXiv {\bf 2010},
arXiv:1004.2515. Available online: https://arxiv.org/abs/1004.2515 (accessed on 20 February 2019).


\bibitem{HSP}
Harder, M.; Salge, C.; Polani D. Bivariate measure of redundant information. {\em Phys. Rev. E} {\bf 2013},  87.
doi:10.1103/PhysRevE.87.012130.

\bibitem{BERT}
Bertschinger, N.; Rauh, J.; Olbrich, E.; Jost, J.; Ay, N. Quantifying Unique Information. {\em Entropy} {\bf 2014}, {\em 16}, 2161.

\bibitem{GK}
Griffith, V.; Koch C.  Quantifying synergistic mutual information. In: {\em Guided self-organization: Inception. Emergence, complexity and computation.} Springer: Berlin/Heidelberg, Germany, 2014, Volume 9, pp.159-190.

\bibitem{RI}
Ince, R.A.A. Measuring multivariate redundant information with pointwise common change in surprisal. {\em Entropy} {\bf 2017}, {\em 19}, 318, doi:10.3390/e19070318.

\bibitem{James}
James, R.G.;  Emenheiser ,J.;  Crutchfield J.P. Unique Information via Dependency Constraints. {\em Journal of Physics A: Mathematical and Theoretical} {\bf 2018}; 52(1):014002.


\bibitem{FL}
Finn, C.; Lizier, J.T. Pointwise Partial Information Decomposition Using the Specificity and Ambiguity Lattices. Entropy {\bf 2018}, 20(4), 297; https://doi.org/10.3390/e20040297.

\bibitem{MGW}
Makkeh, A.; Gutknecht, A.J.; Wibral, M. Introducing a differentiable measure of pointwise shared information. {\em Physical Review E} {\bf 2021}; 103 (3), 032149

\bibitem{IDTxl}
Wollstadt, P.; Lizier, J.T.; Vicente, R.; Finn, C.; Martinez-Zarzuela, M.; Mediano, P.; Novelli, L.; Wibral, M. IDTxl: The Information Dynamics Toolkit xl: a Python package for the efficient analysis of multivariate information dynamics in networks. {\em Journal of Open Source Software} {\bf 2018}; 4(34), 1081. https://doi.org/10.21105/joss.01081.

\bibitem{locinf}
Lizier, J.T. Measuring the Dynamics of Information Processing on a Local Scale. In: {\em Directed Information Measures in Neuroscience}; Wibral, M., Vicente, R., Lizier, J. T. Eds.; Springer: Heidelberg, Germany, 2014,  pp. 161-193.

\bibitem{SBB}
Schneidman, E.;  Bialek, W.;  Berry M. J. Synergy, Redundancy, and Population Codes. {\em J. Neurosci.} {\bf 2003}, 23:11539-11553.

\bibitem{GT}
Gat, I.; Tishby N. Synergy and redundancy among brain cells of behaving monkeys. In: {\em Proceedings of the 1998 conference on Advances in neural information processing systems 2.} Cambridge, MA, USA: MIT Press  1999,   pp. 111-117.

\bibitem{KFP}
 Kay, J.; Floreano D.; Phillips W. A.  Contextually guided unsupervised learning using local multivariate binary processors.  {\em Neural Networks} {\bf 1998}, 11,117-140. 

\bibitem{Bell}
Bell, A. J. The co-information lattice. {\em Proceedings of the Fourth International Symposium on Independent Component Analysis and Blind Signal Separation}. (ICA2003), April 2003, Nara, Japan.

\bibitem{Wibral}
Wibral, M.;  Finn C.; Wollstadt, P.; Lizier, J. T.;  Priesemann, V. Quantifying Information Modification in Developing Neural Networks via Partial Information Decomposition. {\em Entropy} {\bf 2017}, 19, 494.

\bibitem{Conor}
Finn, C.; Lizier, J. T. Quantifying Information Modification in Cellular Automata Using Pointwise Partial Information Decomposition.
{\em Artificial Life Conference Proceedings}, 386-387, 2018. 

\bibitem{Timme}
Timme,  N.; Alford, W.; Flecker, B.;  Beggs J. M. Synergy, redundancy, and multivariate information measures: an experimentalist's perspective. {\em J. Comput. Neurosci.} {\bf 2014}, 36, 119-140. doi:10.1007/s10827-013-0458-4

\bibitem{Timme2}
Sherrill, S. P.;  Timme,;  JM Beggs,;  EL Newman. Partial information decomposition reveals that synergistic neural integration is greater downstream of recurrent information flow in organotypic cortical cultures.
{\em PLoS computational biology} {\bf 2021}, 17 (7), e1009196

\bibitem{Faes}
Pinto, H.; Pernice, R.; Silva, M.E.;  Javorka, M.; Faes, L. Multiscale Partial Information Decomposition of Dynamic Processes with Short and Long-range correlations: Theory and Application to Cardiovascular Control.  arXiv preprint arXiv  {\bf 2022} arxiv.org


\bibitem{Ince1}
Ince, R.A.A.; Giordano, B.L.; Kayser, C.;  Rousselet, G. A.; Gross, J.;  Schyns, P. G. A Statistical Framework
for Neuroimaging Data Analysis Based on Mutual Information Estimated via a Gaussian Copula.
{\em Hum. Brain Mapp.} {\bf 2017}, 38, 1541-1573.

\bibitem{Ince2}
Park, H.;  Ince, R. A. A.;  Schyns, P. G.;  Thut ,G.;   Gross, J.  Representational interactions during audiovisual speech entrainment: Redundancy in left posterior superior temporal gyrus and synergy in left motor cortex. {\em PLoS Biology }{\bf 2018}, 16(8), e2006558. doi:10.1371/journal.pbio.2006558. PMID 30080855

\bibitem{WLP}
Wibral, M.; Lizier, J. T.;  Priesemann, V. Bits from brains for biologically inspired computing. {\em Frontiers in Robotics and  AI} {\bf  2015},  2.  https://doi.org/10.3389/frobt.2015.00005

\bibitem{WPKLP}
Wibral, M.; Priesemann, V.; Kay, J.W.; Lizier, J.T.; Phillips, W.A. Partial information decomposition as a unified approach to the specification of neural goal functions. {\em Brain  Cognit.} {\bf 2017}, {\em 112}, 25-38.

\bibitem{MG}
Graetz, M. Infomorphic Networks: Locally Learning Neural Networks derived from Partial Information Decomposition. Master's Thesis, ETH, Z{\"u}rich, 27-09-2021.

\bibitem{LBJW}
Lizier, J. T.;  Bertschinger, N.;  Jost, J.;  Wibral,  M. Information Decomposition of Target Effects from Multi-Source Interactions: Perspectives on Previous, Current and Future Work. {\em Entropy} {\bf 2018}, 20(4), 307. https://doi.org/10.3390/e20040307


\bibitem{TL} 
Timme, N. M.;  Lapish, C. A Tutorial for Information Theory in Neuroscience. eNeuro, https://doi.org/10.1523/ENEURO.0052-18.2018 

\bibitem{SKBL}
Schulz, J. M.;  Kay, J. W.;  Bischofberger, J.; Larkum, M. E. GABA$_{\rm B}$ Receptor-Mediated Regulation of Dendro-Somatic Synergy in Layer 5 Pyramidal Neurons. {\em Front. Cell. Neurosci.} {\bf 2021},  15:718413. doi: 10.3389/fncel.2021.718413

\bibitem{RM}
Ramaswamy, S.;  Markram, H. Anatomy and physiology of the thick-tufted layer 5 pyramidal neuron. {\em Front. Cell. Neurosci.} {\bf 2015}, doi.org/10.3389/fncel.2015.00233
 
 \bibitem{Schu}
Schuman, B.; Dellal, S.; Pr{\"o}nneke, A.; Machold, R.; Rudy, B.  Neocortical layer 1: an elegant solution to top-down and bottom-up integration. {\em Annual Review of Neuroscience} {\bf 2021}, 44, 221-252. doi.org/10.1146/annurev-neuro-100520-012117
 



\bibitem{ML2013}
Larkum, M.  A cellular mechanism for cortical associations: an organizing principle for the cerebral cortex. {\em Trends Neurosci.} {\bf 2013}, 36,141-151. doi: 10.1016/j.tins.2012.11.00
\bibitem{WS}
Williams S. R.; Stuart, G. J.  Dependence of EPSP efficacy on synapse location in neocortical pyramidal neurons. {\em Science} {\bf 2002}, 295 1907-1910.

\bibitem{BNSPS}
Larkum, M. E.; Nevian, T.; Sandler, M.; Polsky, A.; Schiller J. Synaptic integration in tuft dendrites of layer 5 pyramidal neurons: A new unifying principle. {\em Science} {\bf 2009},  325, 756-760.

\bibitem{FW}
Fletcher, L. N.; Williams, S. R.  Neocortical topology governs the dendritic integrative capacity of layer 5 pyramidal neurons. {\em Neuron} {\bf 2019}, 101, 76-90. https://doi.org/10.1016/j.neuron.2018.10.048

\bibitem{PP}
Poirazi, P.; Papoutsi, A.  Illuminating dendritic function with computational models. {\em Nature Reviews Neuroscience} {\bf 2020}, 21, 303 -321. doi.org/10.1038/s41583-020-0301-7


\bibitem{Shai}
Shai, A.S.; Anastassiou, C. A.; Larkum, M. E.; Koch, C. Physiology of Layer 5 Pyramidal Neurons in Mouse Primary Visual Cortex: Coincidence Detection through Bursting. {\em PLoS Comput. Biol.} {\bf 2015}, 11(3). e1004090. doi:10.1371/journal. pcbi.1004090

\bibitem{HHSMS}
Hay E, Hill S, Schurmann F, Markram H, Segev I. Models of neocortical layer 5b pyramidal cells capturing a wide range of dendritic and perisomatic active properties. {\em PLoS Comput Biol} {\bf  2011},  7. e1002107. doi: 10.1371/journal.pcbi.1002107 PMID: 21829333

\bibitem{Lamme1}
Lamme, V. A. F. Beyond the classical receptive field: Contextual modulation of V1 responses. In  {\em The Visual Neurosciences}.; Werner J. S., Chalupa L. M., Eds. Cambridge, MA: MIT Press, 2004, pp. 720-732. 

\bibitem{Lamme2}
Lamme, V. A. F. Visual Functions Generating Conscious Seeing. {\em Frontiers in Psychology} {\bf 2020}, 11, 83.  10.3389/fpsyg.2020.00083

\bibitem{GS}
Gilbert, C. D.; Sigman, M. Brain states: Top-down influences in sensory processing. {\em Neuron} {\bf 2007},  54, 677-696.

\bibitem{KIDP}
Kay, J. W.;  Ince, R. A. A.;  Dering, B.;  Phillips, W. A. Partial and Entropic Information Decompositions of a Neuronal Modulatory Interaction. {\em Entropy} {\bf 2017}. 19(11), 560.  doi:10.3390/e19110560.

\bibitem{KP2020}
Kay, J. W.;  Phillips, W. A. Contextual Modulation in Mammalian Cortex is Asymmetric. {\em Symmetry} {\bf 2020}, 12, 815.

\bibitem{CT}
Cover, T. M.;  Thomas , J. A. Elements of Information Theory. New York, USA: Wiley-Interscience,  1991.

\bibitem{McG}
McGill, W. J. Multivariate Information Transmission. {\em Psychometrika} {\bf 1954}, 19(2), 97-116.
\bibitem{PICA}
Pica, G.; Piasini, E.; Chicharro, D, ; Panzeri, S. Invariant components of synergy, redundancy, and unique information. {\em Entropy} {\bf 2017}, 19,  451, doi:10.3390/e19090451.



\bibitem{ADM}
Banerjee, P. K.; Rauh, J.;  Montufar, G.  Computing the unique information, In {\em Proceedings of the 2018 IEEE International Symposium on Information Theory}, (Vail, CO), 2018, 141-145.
\bibitem{dit}
James, R. G.; Ellison C. J.; Crutchfield J.P. a Python package for discrete information theory. {\em The Journal of Open Source Software} {\bf 2018}, 25:738. https://doi.org/10.21105/joss.00738. 


\bibitem{R}
R Core Team. R: A language and environment for statistical computing. R
  Foundation for Statistical Computing, Vienna, Austria. URL, 2021.
  https://www.R-project.org/.

\bibitem{reticulate}
 Ushey, K.; Allaire, J. J.; Tang, Y. reticulate: Interface to 'Python'. R package version 1.16, 2020. 
  https://CRAN.R-project.org/package=reticulate
\bibitem{ggplot2}
 H. Wickham. ggplot2: Elegant Graphics for Data Analysis. Springer-Verlag New York, 2016.

\bibitem{coin}
Hothorn, T.; Hornik K.;  van de Wiel M. A.; Zeileis A. Implementing a class of permutation tests: The coin package
{\em Journal of Statistical Software} {\bf 2008}, 28(8), 1-23. doi: 10.18637/jss.v028.i08 (URL: https://doi.org/10.18637/jss.v028.i08).
\bibitem{Palmer}
Palmer, L. M.;  Schulz, J. M.; Murphy, S. C.; Ledergerber, D.; Murayama, M.; Larkum, M. E. The cellular basis of GABA(B)-mediated interhemispheric inhibition. {\em Science} {\bf 2012}  335, 989-993. doi: 10.1126/science.1217276






\bibitem{Shai-dat}
Available online: 
https://senselab.med.yale.edu/ModelDB/ShowModel.cshtml?model=180373\&\\file=/ShaiEtAl2015/data/spikes\_.dat\#tabs-2 (accessed on 12 June  2022).


\end{thebibliography}
\end{document}